\documentclass[times]{elsarticle}

\usepackage{amssymb}
\usepackage{comment}
\usepackage{amsmath}
\usepackage{amsthm}
\usepackage{algorithmic}
\usepackage{moreverb} 
\usepackage[colorlinks,bookmarksopen,bookmarksnumbered,citecolor=Equationlor=red,pdfauthor=author]{hyperref}
\usepackage{amsfonts,amsmath,amssymb}
\usepackage{psfrag,pstricks,pst-node}

\usepackage[section]{placeins}
\usepackage{subfigure} 
\usepackage{mathrsfs}
\usepackage{graphicx}
 \usepackage[ruled,vlined]{algorithm2e}
\usepackage{fancyhdr}
\usepackage{psfrag} 
\usepackage[export]{adjustbox} 
\usepackage{enumerate} 
\usepackage{comment}
\usepackage[margin=2.5cm]{geometry}
\usepackage{ulem} 
\usepackage{multirow}
\usepackage{makecell}

\newcommand{\bmX}{\mathbf X}

\newcommand{\bmu}{\mathbf u}

\newcommand{\bmw}{\mathbf w}
\newcommand{\bmx}{\mathbf x}
\newcommand{\bmy}{\mathbf y}

\newcommand{\vect}[1]{\boldsymbol{#1}}
\newcommand{\rv}[1]{\lvert #1\rangle}

\newcommand{\me}[3]{\langle #1\lvert #2 \rvert #3 \rangle}

\begin{document}

\begin{frontmatter}
\title{Quantum machine learning for efficient reduced order modelling of turbulent flows}
\author[1]{Han Li}
\ead{2511903@tongji.edu.cn}
\author[1]{Yutong Lou}
\ead{lyt@tongji.edu.cn}
\author[1]{Dunhui Xiao\corref{cor1}}
\ead{xiaodunhui@tongji.edu.cn}
\cortext[cor1]{Corresponding author}
\affiliation[1]{organization={School of Mathematical Sciences, Key Laboratory of Intelligent Computing and Applications(Ministry of Education), Tongji University},
            addressline={NO.1239 Siping Rd.}, 
            city={Shanghai},
            postcode={200092}, 
           country={CHINA}}

\begin{abstract}
Accurately predicting turbulent flows remains a central challenge in fluid dynamics due to their high dimensionality and intrinsic nonlinearity. Recent developments in quantum algorithms and machine learning offer new opportunities for overcoming the computational barriers inherent in turbulence modeling. Here we present a new hybrid quantum-classical framework that enables faster-than-real-time turbulence prediction by integrating machine learning, quantum computation, and fluid dynamics modeling, in particular, the reduced-order modeling. The novel framework combines quantum proper orthogonal decomposition (QPOD) with a quantum-enhanced deep kernel learning (QDKL) approach. QPOD employs quantum circuits to perform efficient eigenvalue decomposition for low-rank flow reconstruction, while QDKL exploits quantum entanglement and nonlinear mappings to enhance kernel expressivity and dynamic prediction accuracy. The new method is demonstrated on three benchmark turbulent flows, our architecture achieves significantly improved predictive accuracy at reduced model ranks, with training speeds up to 10 times faster and parameter counts reduced by a factor of $1/N$ compared to classical counterparts, where N is the input dimensionality. Although constrained by current noisy intermediate-scale quantum (NISQ) hardware, our results demonstrate the potential of quantum machine learning to transform turbulence simulation and lay a solid foundation for scalable, real-time quantum fluid modeling in future quantum computers.
\end{abstract}

\end{frontmatter}

\section{Introduction}\label{sec:introduction}

Turbulence is a pervasive phenomenon in nature and finds extensive applications across scientific and engineering disciplines, ranging from aircraft design to biomedical engineering. Understanding turbulence is critical to advancing many scientific and technological fields. Simulating turbulent flows poses significant challenges due to the high-dimensional nonlinear dynamics and the exponential scaling of energy-containing eddies with the high Reynolds number, which create prohibitive computational barriers for direct numerical simulation (DNS) \cite{Pope_2000, Holmes2012, GIVI2021}. Accordingly, the development of reduced-order models (ROMs) for fluid dynamics has garnered increasing attention, as it facilitates the reduction of problem dimensionality while efficiently capturing the intrinsic dynamical characteristics of flow \cite{Taira2017}. The importance of developing ROMs lies in their ability to facilitate the identification of appropriate low-dimensional representations, which can then be utilized to reduce the computational costs of numerical simulations \cite{Brunton2020}. Traditional intrusive ROMs necessitates deep modifications to solver internals, obstructing integration with commercial computational fluid dynamics (CFD) workflows, which leads to  significant limitations \cite{chen2012blackbox}. As a result, non-intrusive ROMs (NIROMs) leveraging machine learning have advanced specific applications. Among all commonly used model reduction techniques, proper orthogonal decomposition (POD) stands as a cornerstone methodology for capturing dominant coherent structures through modal energy decomposition\cite{Berkooz1993}. The stacked autoencoders (SAE) coupled with dynamic mode decomposition (DMD) enable efficient pier flow prediction, while convolutional autoencoders (CAE) achieve multiphysics field reconstruction \cite{Noack2011,Noori2013,ZHU2024118308,GAO2025112517}. Additionally, methodologies employing singular value decomposition (SVD) to generate POD basis functions with Gaussian process regression (GPR) predicting POD coefficient evolution were established\cite{XIAO2019323}. However, these approaches persistently fail to capture essential turbulence physics with three critical limitations remaining unresolved: they cannot adequately resolve Kolmogorov-scale phenomena and inertial-range intermittency; they lack capacity to model transient vortex shedding, rupture, and separation bubble dynamics; and they poorly characterize parameter-dependent bifurcations during flow separation\cite{Taira2017, Brunton2020, Sagaut2006, Rowley2017, Peter2015, Peherstorfer2020}. These shortcomings stem intrinsically from the $O(N^3)$ complexity of classical SVD for basis construction and the restricted expressivity of classical kernels in modeling turbulent energy transfer. As fluid systems scale toward industrial complexity, this impasse demands a computational paradigm shift to achieve high-fidelity simulations capable of resolving nonlinear multiscale features with tractable costs.

Remarkably, quantum computing offers a transformative pathway, where its inherent parallelism and high-dimensional state representation align fundamentally with turbulence physics. Quantum machine learning (QML) synergizes quantum algorithms with classical learning to enable exponential accelerations in data processing, as demonstrated by pioneering frameworks including quantum neural networks (QNNs) for fluid systems, quantum Fourier neural operators for PDE learning, and quantum reservoir computing\cite{Biamonte2017, Gupta2001, Yadav2023, Markidis2022, Xiao2024, Jain2024, Mujal2021, Wei2023}. Recently, quantum-enhanced ROM components are emerging\cite{Asztalos2024}. Quantum principal component analysis (QPCA) achieves exponential speedup in basis generation through quantum unitary transformations\cite{Lloyd2014}. Resource-efficient variants like resonant QPCA and variational quantum SVD (VQSVD) specifically address noisy intermediate-scale quantum (NISQ) constraints\cite{LiQPCA2021,Wang2021}. Meanwhile, quantum regression techniques accelerate predictive modeling, including quantum-assisted Gaussian process regression (GPR) and low-rank Hilbert space methods\cite{zhao2019,Farooq2024}. Despite these advances, a unifying framework capable of overcoming the fundamental efficiency bottlenecks in turbulence ROMs, specifically the extensive training time and massive parameter scales required for predictive accuracy, has remained out of reach.

Here we pioneer a hybrid quantum-classical ROM framework that integrates quantum proper orthogonal decomposition (QPOD) for accelerated basis construction with quantum deep kernel learning (QDKL) for turbulence prediction, specifically engineered to overcome classical bottlenecks. QPOD exploits quantum unitary transformations to generate high-fidelity orthogonal POD bases with minimal parameters, achieving a $10\times$ reduction in training time and slashing parameter complexity to $1/N$ of classical POD-DKL implementations. Simultaneously, QDKL fuses quantum circuits with deep kernels to leverage entanglement and nonlinear gates, explicitly encoding turbulent complexity through Matérn-kernel-enhanced quantum feature maps that capture vortex stretching and tilting dynamics. Our architecture integrates two synergistic innovations: quantum-assisted spectral decomposition of Reynolds stress tensors via parameterized unitary circuits and a hybrid optimizer combining classical gradient descent with quantum approximate optimization for robust time-evolution prediction. This combined approach unlocks unprecedented capability in capturing high-dimensional nonlinearities and transient phenomena. While current NISQ hardware noise limits small-scale mode resolution, our framework establishes the computational blueprint for the fault-tolerant quantum era—poised to surpass classical ROMs in speed, accuracy, and scalability as hardware matures, ultimately enabling real-time, high-fidelity turbulence simulation.

\section{Result}\label{sec:result}

\subsection{VQAs as the basis for quantum-enhanced NIROM framework}
Variational quantum algorithms (VQAs) are currently one of the most practical and promising research directions in quantum computing, especially on NISQ devices. A significant importance of VQAs lies in providing a universal framework for solving optimization problems\cite{Cerezo2021}. Thus, we employ variational quantum algorithms to establish a quantum-enhanced NIROM framework with quantum advantages.

Considering a mathematical problem to be solved, the first step to develop a VQA is to design a loss function $L$ so that it can be transformed into an optimization problem whose solution helps solve the original problem. To address the optimization problem $\vect{\theta}^*=\arg\min_{\vect{\theta}} L(\vect{\theta})$, an ansatz is to be proposed, that is, a unitary operator controlled by a set of parameters $\vect{\theta}$, which can be trained by a classical optimizer, forming a hybrid quantum-classical training loop. It will be shown how the VQAs can provide us with an opportunity to harness the quantum advantage in the field of ROM.

We aim to construct a NIROM that utilizes POD for model reduction and employs DKL for prediction. However, classical POD extracts dominant modes based on SVD. Although mathematically rigorous, this method struggles to overcome the computational difficulties associated with large-scale problems, which the inherent parallelism of quantum computing effectively addresses. Furthermore, classical DKL are fundamentally constrained by the computational costs of nonlinear feature extraction and hyperparameter optimization. Utilizing the ansatz circuit from VQAs as a feature extractor could potentially bring significant optimization to the algorithm.

We first improve the POD method by applying VQAs. The core step of POD involves performing SVD on a specific matrix to obtain its left singular vectors and singular values. For a matrix $S$, achieving the above objective only requires solving the eigenproblem,
\begin{equation}
    S^TSu_i = \lambda_iu_i.
\end{equation}
The computational complexity of solving this eigenproblem reaches $O(N^3)$, posing significant challenges for classical computing when handling large-scale matrices. It is encouraging that after applying the VQAs, the time complexity of the ansatz circuit can achieve $O(N^2)$ or even lower. Suppose $M = S^T S$. Clearly, solving the aforementioned eigenproblem is equivalent to searching for a set of orthonormal bases $\{u_j\}$ in the linear space such that $\hat{\lambda}_j = u_j^T M u_j$ exactly equals the eigenvalue $\lambda_j$ of $M$, meaning the reconstructed matrix $\hat{M} = \sum_j \hat{\lambda}_j u_j u_j^T$ precisely equals the matrix $M$. To search for such a set of orthonormal bases, we can utilize the property that unitary transformations preserve lengths and orthogonality. Suppose that $\{\rv{\psi_1},\cdots,\rv{\psi_n}\}$ is a standard orthogonal basis of $\mathbb{R}^n$, obviously the result of the unitary transformation $U$ acting on this basis is also a standard orthogonal basis, which is noted as $\{\rv{u_1},\cdots,\rv{u_n}\}$ with $\rv{u_j}=U\rv{\psi_j},j=1,2,\cdots,n$. We use an ansatz circuit $U(\vect{\theta})$ to learn the unitary transformation such that $\rv{u_j}$ is the eigenvector of $M$, where $\vect{\theta}$ represents the full set of parameters for the parameterized quantum circuit. As a symmetric matrix, $M$ can be decomposed into a linear combination of Pauli terms, i.e. $M=\sum c_kA_k$, to estimate the expectation value
$\langle M\rangle_j=\me{u_j}{M}{u_j}$ via Hadamard test on $\rv{u_j}$. Design the loss function
\begin{equation}
    \mathcal{L}(\vect{\theta})=\left\Vert M-\sum_{j=1}^n \langle M\rangle_j\left|u_j\right\rangle\left\langle u_j\right|\right\Vert_2
\end{equation}
Minimize the loss function $\mathcal{L}$, and set $\vect{\theta}^*=\arg\min_{\vect{\theta}} \mathcal{L}(\vect{\theta})$. Then $\langle M\rangle_j=\me{\psi_j}{U(\vect{\theta^*})^\dagger MU(\vect{\theta^*})}{\psi_j}$ is the $j$th eigenvalue of $M$, and $\rv{u_j}=U(\vect{\theta^*})\rv{\psi_j}$ is the corresponding eigenvector.

\begin{algorithm}\label{alg:qsvd}
    \SetAlgoLined
    \KwIn{symmetric positive definite matrix $M=\sum c_kA_k\in\mathbb{R}^{n\times n}$, parametrized circuits $U(\vect{\theta})$ with initial parameters $\vect{\theta}_0$, a standard orthogonal basis $\{\rv{\psi_j}\}_{j=1}^n$;}
    \For{$i=1,2,\cdots,\mathrm{itermax}$}{
        \For{$j=1,2,\cdots,n$}{
        Apply $U(\vect{\theta})$ to state $\rv{\psi_j}$ and define $\rv{u_j}=U(\vect{\theta})\rv{\psi_j}$\;
        Compute $\langle M\rangle_j=\me{u_j}{M}{u_j}$ via Hadamard tests\;
    }
    Compute the loss function $\mathcal{L}(\vect{\theta})=\left\Vert M-\sum_{j=1}^n \langle M\rangle_j\left|u_j\right\rangle\left\langle u_j\right|\right\Vert_2$\;
    Perform optimization to minimize $\mathcal{L}(\vect{\theta})$, update $\vect{\theta}$\;
    \If{$\mathcal{L}(\vect{\theta})$ converges with tolerance $\varepsilon$}{break\;}
    }
    \KwOut{eigenvalues $\{\langle M\rangle_j\}_{j=1}^n$, eigenvectors $\left\{U(\vect{\theta})\rv{\psi_j}\right\}_{i=1}^n$}
\caption{Variational Quantum Eigensolving}\label{alg1}
\end{algorithm}

Furthermore, we leverage VQAs to improve the feature extractor in DKL. Let us start from classical DKLframework. Classical DKL combines deep neural networks (DNNs) with GPR, enhances generalization by leveraging DNNs’ nonlinear feature extraction and GPR’s small-sample learning \cite{wilson16}. Select a base kernel function $k_0(\bmx,\bmx';\vect{\theta})$ with hyperparameters $\vect{\theta}$. In the DKL method, a DNN is used as a feature extractor to map the data $\bmx$ to $\phi(\bmx;\bmw)$, based on which we can further construct a learnable kernel function $k(\bmx,\bmx')=k_0(\phi(\bmx;\bmw),\phi(\bmx';\bmw))$. Applying GP on $\mathcal{D}'=\{(\phi(\bmx_i),y_i)\}_{i=1}^n$ gives the log marginal likelihood function $\mathcal{L}$ of the targets $\bmy$, and
\begin{equation}\label{logmar}
    \mathcal{L}(\bmw,\vect{\theta};\bmy,\bmX)=\log(p(\bmy\vert\bmw,\vect{\theta},\bmX)) \propto -\left(\bmy^T(K_{\bmw,\vect{\theta}}+\sigma^2I)^{-1}\bmy+\log|K_{\bmw,\vect{\theta}}+\sigma^2I|\right),
\end{equation}
where $K_{\bmw,\vect{\theta}}$ notes $k_0(\phi(\bmX;\bmw),\phi(\bmX;\bmw);\vect{\theta})$, and $\sigma$ is the standard deviation of observation noise. The parameters $\bmw$ in the DNN together with the hyperparameters $\vect{\theta}$ of the base kernel function form the parameters $\vect{\gamma}=\{\bmw,\vect{\theta}\}$ to be learned in the DKL, by maximizing the logarithmic marginal likelihood $\mathcal{L}(\vect{\gamma};\bmy,\bmX)$ in Eq.~\ref{logmar} of the Gaussian process.

\begin{figure}[ht]
\centering
\includegraphics[width = 1.0\linewidth,angle=0,clip=true]{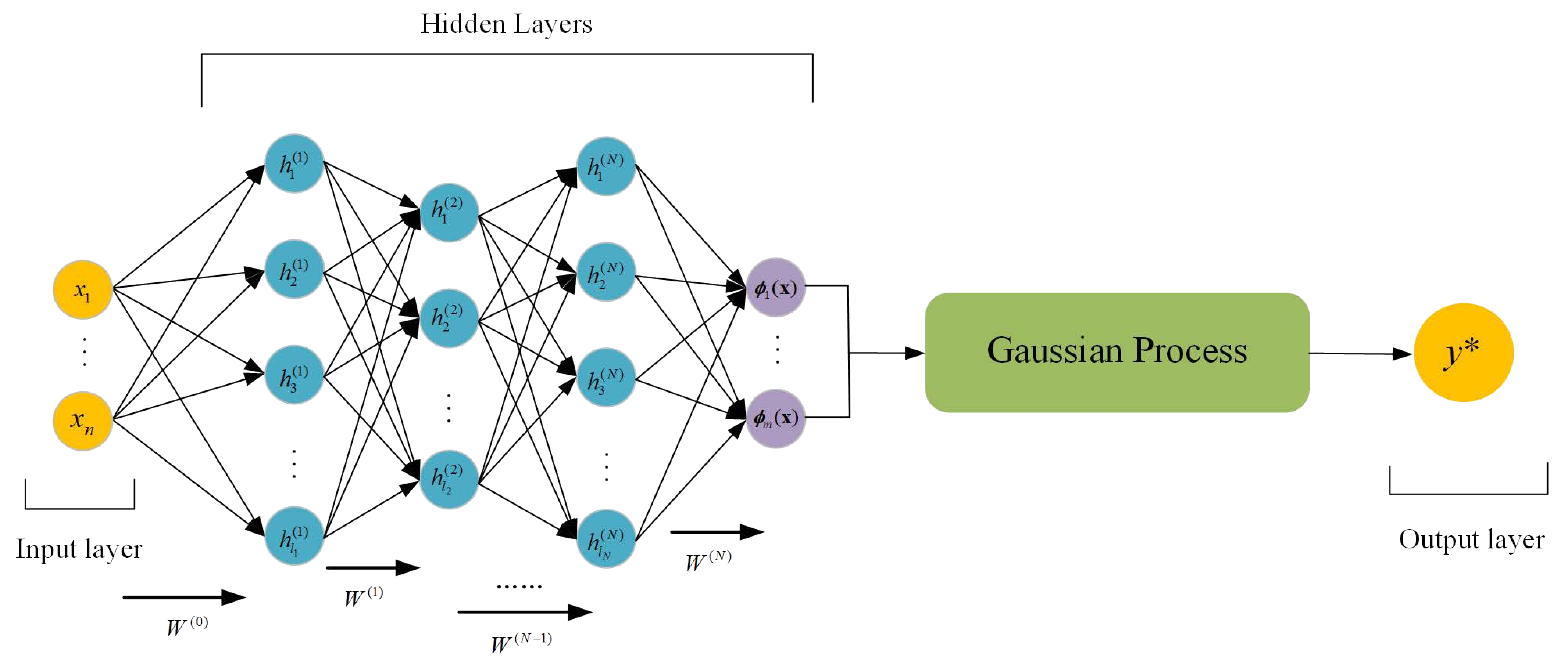} 
\caption{The structure of the deep kernel learning.}
\label{fig:dpl}
\end{figure}

However, classical feature extractors still have their limitations in the feature extraction of turbulent flows.
To overcome these limitations, we propose a hybrid quantum-classical DKL (QDKL) architecture that synergizes classical and quantum processing. The classical preprocessing network transforms input data into quantum state space representations. Through activation function (e.g., sigmoid/tanh) modulation of angular parameters that restricts the angular domain to $[0, \pi/2]$, classical information is encoded into rotational angles of RY (rotation Y) gates, which are subsequently introduced into quantum circuits (ansatz) for nonlinear transformations via coherent superposition and entanglement operations. In the process of feature extraction, the classical network is responsible for data degradation and feature smoothing, and the quantum circuit provides high-dimensional nonlinear feature space, which complement each other to enhance the model generalization ability. The output of the quantum feature extractor is used as the input of the GP, and the Matern kernel ($\nu$=2.5) is used to construct the deep kernel function, which provides a better match to the local curvature of the Bloch sphere manifold onto which the data are typically transformed by quantum embeddings (e.g. AngleEmbedding). Finally, the parameters in the NN and ansatz together with the hyperparameters of the Matern kernel are to be learned by maximizing the logarithmic marginal likelihood of the GP.

\begin{figure}[ht]
\centering
\includegraphics[width = 0.9\linewidth,angle=0,clip=true]{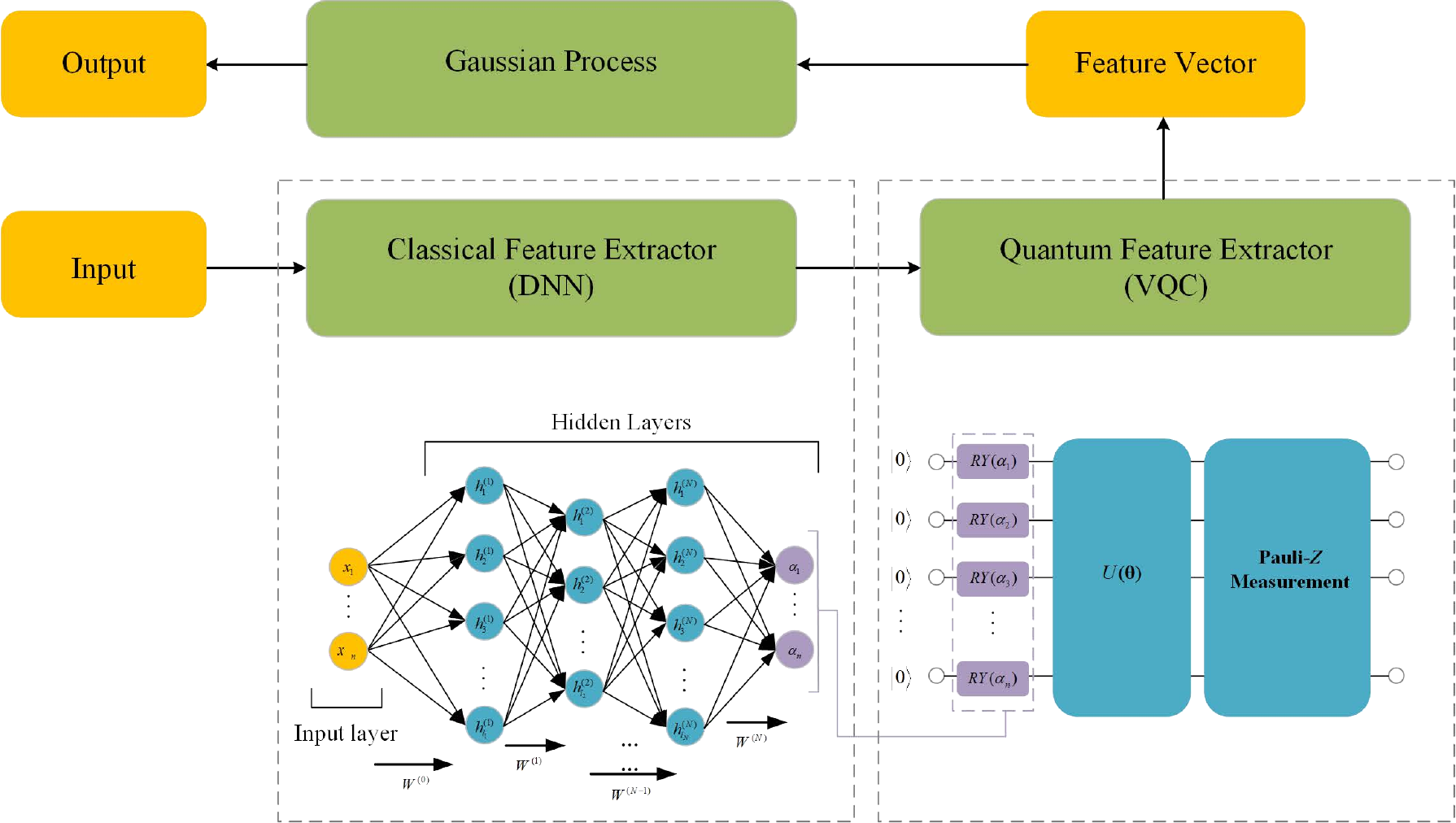}
\caption{The structure of the hybrid quantum-classical deep kernel learning.}\label{fig:qdpl}
\end{figure}

\subsection{Construct quantum-enhanced NIROM framework}

The physics of turbulent flows is fundamentally governed by the incompressible Navier-Stokes equations, which encapsulate mass and momentum conservation:
\begin{equation}
    \begin{aligned}
        &\nabla\cdot\mathbf{u}=0, \\
        &\frac{\partial\mathbf{u}}{\partial t} + \mathbf{u}\cdot\nabla\mathbf{u} = -\nabla p + \nu\nabla^2\mathbf{u},
    \end{aligned}
    \label{eq:NS}
\end{equation}
where $\mathbf{u}=(u,v)^T$ denotes the velocity field, $p=\tilde{p}/\rho_0$ the normalized pressure, and $\nu$ the kinematic viscosity. The nonlinear advection term $\mathbf{u}\cdot\nabla\mathbf{u}$ constitutes the primary source of turbulence complexity, generating multiscale interactions that challenge conventional reduced-order modeling approaches.

For computational implementation, we discretize Eq.~\eqref{eq:NS} in space to obtain the high-fidelity full-order model (FOM):
\begin{equation}
    \begin{aligned}
        &\mathbf{C}^T\mathbf{u} = 0 \\
        &\mathbf{M}\frac{d\mathbf{u}}{dt} = \mathbf{A}(\mathbf{u})\mathbf{u} + \mathbf{K}\mathbf{u} + \mathbf{C}\mathbf{p}
    \end{aligned}
    \label{eq:disc_NS}
\end{equation}
Here, $\mathbf{u}$ and $\mathbf{p}$ represent the discrete velocity and pressure vectors at the nodal points. The key operators involved are: the mass matrix $\mathbf{M}$, the diffusion matrix $\mathbf{K}$, the pressure gradient operator $\mathbf{C}$, and critically the solution-dependent nonlinear advection operator $\mathbf{A}(\mathbf{u})$. This velocity-dependent nonlinear term dominates the turbulent energy cascade and presents the central challenge for model reduction.

We initiate quantum-enhanced POD by defining the velocity snapshot matrix:
\begin{equation}
    S=\begin{bmatrix}
        \mathbf{u}^1 & \mathbf{u}^2 & \cdots & \mathbf{u}^{N_t}
    \end{bmatrix}
\end{equation}
where $N_t$ snapshots capture the flow evolution, with each $\mathbf{u}^n = [u_1^n, \cdots, u_N^n, v_1^n, \cdots, v_N^n]^T$ containing velocity components at $N$ spatial nodes. Apply the variational quantum eigensolver on the eigenvalue problem:
\begin{equation}
    S^TS\hat{u}_i = \lambda_i\hat{u}_i
\end{equation}.
Then we can obtain POD basis $\Phi = [\boldsymbol{\phi}_1, \cdots, \boldsymbol{\phi}_m]$ through energy thresholding, where $\vect{\phi}_j=S\hat{u}_j$. For a tolerance $\eta<1$, choose the first $m$ left singular vectors as POD basis functions, where $m$ satisfies $\sum_{i=1}^m\sigma_i^2/\sum_{i=1}^{N_t}\sigma_i^2\ge \eta,m\le N_t,\sigma_i=\sqrt{\lambda_i}$. For any variable $\bmu$, it can be expressed as
\begin{equation}
    \bmu=\bar{\bmu}+\sum_{j=1}^m\alpha_j\vect{\phi}_j,
\end{equation}
where $\alpha_j=\langle\bmu-\bar{\bmu},\vect{\phi}\rangle$ and $\bar{\bmu}$ are the means of the variable $\bmu$.

After obtaining the POD basis functions through the QPOD method, we implement the approximation of governing equations by training a QDKL network. By projecting all snapshots onto the reduced-order space, the coefficients corresponding to the POD basis functions can be derived. The POD coefficients at the $k$th time level are noted as
\begin{equation}
    \vect{\alpha}^k=(\alpha_1^k,\alpha_2^k\cdots,\alpha_m^k).
\end{equation}
By training the QDKL network, we obtain a function $f_j$ corresponding to each POD basis function, which maps the POD coefficients at one time level $\vect{\alpha}^{k-1}$ to a component $\alpha_j^k$ of the POD coefficients at the subsequent time level, i.e.
\begin{equation}
    f_j:\vect{\alpha}^{k-1}=(\alpha_1^{k-1},\alpha_2^{k-1}\cdots,\alpha_m^{k-1})\mapsto \alpha_j^k
\end{equation}

This completes the construction of the entire NIROM, integrating QPOD and QDKL architectures. The framework has been validated through numerical experiments on canonical fluid flow benchmarks, demonstrating its capability to capture nonlinear dynamics while maintaining computational tractability.

\begin{figure}[ht]
\centering
\includegraphics[width = \linewidth,angle=0,clip=true]{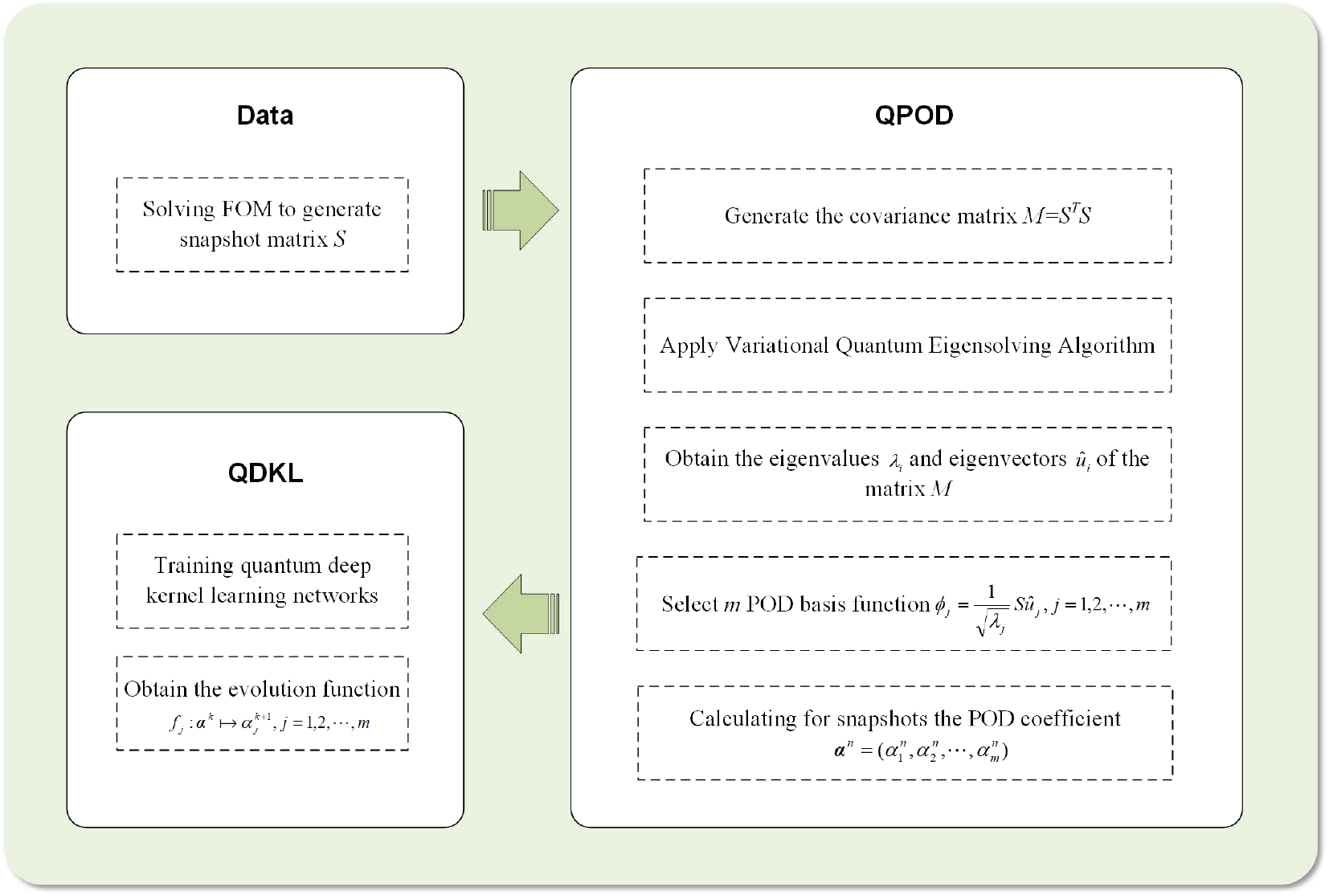}
\caption{The structure of the NIROM.}\label{fig:nirom}
\end{figure}

\subsection{Advantages from an error perspective}
Following the construction of a quantum-enhanced reduced-order model, a quantitative analysis of its approximation error is of significant interest. This analysis aims to elucidate the potential advantages of variational quantum circuits (VQCs) over classical neural networks (NNs) within the framework of deep kernel learning (DKL) from an error perspective.

We begin by considering a standard GPR. For a positive definite kernel \( k: \mathcal{X} \times \mathcal{X} \to \mathbb{R} \), there exists a unique reproducing kernel Hilbert space (RKHS), denoted by \( \mathcal{H}_k \), which is a subspace of \( \mathbb{R}^{\mathcal{X}} \) equipped with an inner product satisfying the reproducing property\cite{Aronszajn1950,Berlinet2004}:
\begin{equation}
    \langle k(\cdot, x), k(\cdot, y) \rangle_{\mathcal{H}_k} = k(x, y).
\end{equation}
To simplify the analysis, we assume a noiseless dataset \( \mathcal{D} = \{(x_i, y_i)\}_{i=1}^n \) where \( y_i = f(x_i) \) for an underlying function \( f \in \mathcal{H}_k \). In this setting, GP regression is equivalent to the following RKHS interpolation problem:
\begin{align}
\min_{s \in \mathcal{H}_k} \quad & \Vert s \Vert_{\mathcal{H}_k} \\
\mathrm{s.t.} \quad & s(x_i) = f(x_i), \quad \forall i = 1, \dots, n.
\end{align}
The solution to this problem is the minimum-norm interpolant. A key result in RKHS theory provides a pointwise error estimate for this interpolant:
\begin{equation}
    |s(x) - f(x)| \leq \Vert f \Vert_{\mathcal{H}_k} \cdot P_X(x),
\end{equation}
where \( P_X(x) \) is the power function\cite{Robert1995,Wendland2004}. It is defined as
\begin{equation}
    P_X(x) = \sqrt{ k(x, x) - K_X(x) K^{-1} K_X(x)^T }.
\end{equation}
Here, \( K_X(x) = [k(x, x_1), \cdots, k(x, x_n)] \) is the vector of kernel evaluations between \( x \) and the data points, and \( K \) is the Gram matrix with entries \( K_{ij} = k(x_i, x_j) \).

Now let us define \( g(x) = K_X(x) K^{-1} K_X(x)^T \). The power function can then be expressed as \( P_X(x) = \sqrt{k(x, x) - g(x)} \). To understand the local behavior of the error, consider a point \( x = x_p + \delta \), where \( x_p \) is the closest data point in the set \( X \). A Taylor expansion of \( g(x) \) around \( x_p \) yields:
\begin{equation}
    g(x) = g(x_p) + \delta^T \nabla g(x_p) + \frac{1}{2} \delta^T \, \text{Hess}\, g(x_p) \, \delta + o(\Vert \delta \Vert^2).
\end{equation}

We focus on the Matérn kernel with smoothness parameter \( \nu = 2.5 \), a common choice that offers a balance between smoothness and flexibility. Its form is:
\begin{equation}
    k(x,y) = \sigma^2 \left( 1 + cr + \frac{c^2r^2}{3} \right) \exp(-cr),
\end{equation}
where \( c = \sqrt{5}/\ell \), \( \ell \) is the length-scale parameter, and \( r = \Vert x - y \Vert \) is the Euclidean distance.

By substituting this specific kernel into the Taylor expansion and leveraging the fact that \( g(x_p) = k(x_p, x_p) \) and \( \nabla g(x_p) = 0 \) (due to the stationarity of the kernel and the interpolating condition), we derive:
\begin{equation}
    g(x) = k(x_p, x_p) - \frac{c^2 \sigma^2}{3} \Vert \delta \Vert^2 + o(\Vert \delta \Vert^2).
\end{equation}
Consequently, the power function simplifies to:
\begin{equation}
    P_X(x) = \sqrt{k(x, x) - g(x)} = C \Vert \delta \Vert + o(\Vert \delta \Vert),
\end{equation}
where \( C = c \sigma/\sqrt{3}\). This establishes that the local error is linearly bounded by the distance to the nearest data point.

In deep kernel learning, the base kernel is applied to features transformed by a parametric model, i.e., \( k_{\text{DKL}}(x, y) = k(\phi(x), \phi(y)) \), where \( \phi : \mathbb{R}^{m_1} \to \mathbb{R}^{m_2} \) is a non-linear mapping (e.g., a neural network or a quantum circuit). The corresponding power function becomes:
\begin{equation}
    P_{\phi, X}(x) = C \Vert \delta_\phi \Vert + o(\Vert \delta_\phi \Vert), \quad \text{with} \quad \delta_\phi = \phi(x) - \phi(x_p).
\end{equation}
Assuming \( \phi \) is differentiable, we can perform a first-order expansion: \( \delta_\phi \approx J_\phi(x_p) \, \delta \), where \( J_\phi \) is the Jacobian matrix of \( \phi \) at \( x_p \). This leads to:
\begin{equation}
    P_{\phi, X}(x) \approx C \Vert J_\phi \, \delta \Vert + o(\Vert \delta \Vert) \leq C \Vert J_\phi \Vert_F \Vert \delta \Vert + o(\Vert \delta \Vert),
\end{equation}
where \( \Vert \cdot \Vert_F \) denotes the Frobenius norm. The final error estimate for the deep kernel is therefore:
\begin{equation}
    |s(x) - f(x)| \leq C \Vert f \Vert_{\mathcal{H}_k} \cdot \Vert J_\phi \Vert_F \cdot \Vert \delta \Vert + o(\Vert \delta \Vert).
\end{equation}
This result highlights that the model's generalization error is influenced not only by the data density (\( \Vert \delta \Vert \)) but also by the sensitivity of the feature map, quantified by \( \Vert J_\phi \Vert_F \).

The critical difference between classical and quantum approaches lies in the fundamental behavior of the Frobenius norm of the feature map Jacobian $\Vert J_\phi \Vert_F $. For a classical NN, where the feature map $\phi$ is constructed through the multi-layered composition of weight matrices and non-linear activation functions such as ReLU or Tanh, the Jacobian norm $\Vert J_\phi \Vert_F$ is in principle unbounded. The parameters of the network reside in an unconstrained Euclidean space, and their norms can grow arbitrarily large during training. Consequently, the magnitude of $\Vert J_\phi \Vert_F$ becomes highly dependent on both the input data and the optimized weights, leading to potentially large and unstable error bounds. 

In contrast, feature maps constructed via parameterized quantum circuits exhibit fundamentally different mathematical properties. Consider a variational quantum circuit employing angle encoding $U(x)=\bigotimes_{i=1}^{m_1}R_y(x_i)$, followed by a parameterized circuit $V(\theta)$. The feature map is defined as the expectation value of an observable $O$:
\begin{equation}
    \phi_j(x) = \langle 0 | U^\dagger(x) V^\dagger(\theta) O_j V(\theta) U(x) | 0 \rangle,
\end{equation}
where \( \phi(x) \in \mathbb{R}^{m_2} \). The output of this architecture is inherently bounded by the spectral norm of the observable $O_j$, such that $|\phi_j(x)| \leq \Vert O_j \Vert$, a direct consequence of the normalization of quantum states and the norm-preserving nature of unitary evolution.
More profoundly, the elements of the Jacobian matrix $\partial \phi_j / \partial x_i$ can be computed exactly using the parameter-shift rule. This rule dictates that for parameters encoded with Hermitian generators, as is the case with $R_y$ rotation gates, the partial derivative is given exactly by the difference of two expectation values, that is
\begin{equation}
    \frac{\partial \phi_j}{\partial x_i} = \frac{1}{2} \left(\phi_j\left(x_i + \frac{\pi}{2}\right) - \phi_j\left(x_i -\frac{\pi}{2}\right)\right).
\end{equation}
Since each expectation value on the right-hand side is individually bounded, their difference is necessarily bounded by a constant $\Gamma$, typically on the order of the observable's spectral norm. It follows that every element of the Jacobian matrix satisfies $|[J_\phi]_{ji}| \leq \Gamma$, which immediately implies a strict upper bound on the Frobenius norm: 
\begin{equation}
    \Vert J_\phi \Vert_F \leq \sqrt{ \sum_{i=1}^{m_1} \sum_{j=1}^{m_2} |[J_\phi]_{ji}|^2 } \leq \sqrt{m_1 m_2} \, \Gamma.
\end{equation}
This upper bound depends only on the input and output dimensions and is independent of the specific values of the circuit parameters $\theta$.

This mathematical property signifies that the quantum feature map $\phi$ is Lipschitz continuous with an a priori known and controllable Lipschitz constant. From a machine learning perspective, this has profound implications. It leads to tighter and more stable theoretical bounds on generalization error, enhancing the predictability of model behavior on unseen data. Furthermore, it imbues the model with an inherent robustness against adversarial perturbations, as small changes in the input cannot induce large changes in the output. When integrated into hybrid quantum-classical networks, the stability of the quantum feature map can also contribute to more stable training dynamics by preventing pathological gradient behavior.

The analysis demonstrates that the error bound for a VQC-based deep kernel is explicitly controlled by \( \Vert J_\phi \Vert_F \leq \sqrt{m_1 m_2} \, \Gamma \). In contrast, the analogous bound for a classical NN lacks such a guaranteed upper limit. Therefore, deep kernel learning models based on variational quantum circuits are expected to exhibit more stable and potentially faster convergence to the underlying true function, as their worst-case error is constrained by the inherent mathematical properties of quantum expectation values. This theoretical advantage stems from the bounded nature of the quantum feature map's Jacobian, a property not generally shared by classical neural networks.

\begin{figure}[htbp!]
\centering
\includegraphics[width = 0.6\linewidth,angle=0,clip=true]{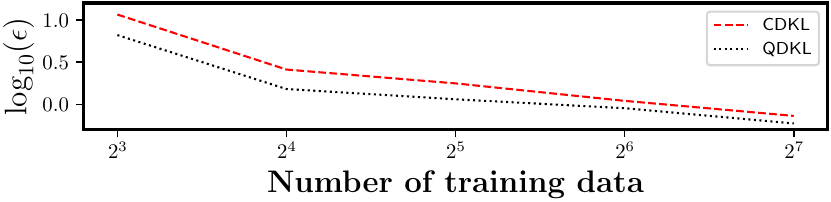}  
\caption{The error on the test set using classical deep kernel learning model and quantum-enhanced deep kernel learning model.}
\label{fig:dkl_convergence}
\end{figure}

To validate the theoretical conclusions above, we designed a numerical experiment. We selected the target function $f(x) = \sin x + 0.1x^2$ and sequentially sampled $2^3, 2^4, 2^5, 2^6, 2^7$ equidistant training points with added Gaussian noise (standard deviation $\sigma=0.2$) over the interval $[-2\pi, 2\pi]$. We compared a classical deep kernel learning model and a quantum-enhanced deep kernel learning model, both with a parameter size of 16. Each model was trained for 300 epochs with a learning rate of 0.02. The model error on the test set was defined as \( \epsilon = \Vert f_{\text{model}} - f_{\text{true}} \Vert_2 \). The experimental results are shown in the Figure \ref{fig:dkl_convergence}. As can be seen from the figure, the quantum-enhanced method consistently yields significantly smaller errors on the test set as the number of training points increases, which clearly demonstrates its superiority.

\subsection{Application to turbulent flows}

In the first numerical example, the classical fluid mechanics problem of the lid-driven cavity flow with Reynolds number $Re = 4000$ is simulated. For this case, the computational domain is defined as a two-dimensional square cavity $ \Omega = [0,1] \times [0,1] $, governed by the incompressible Navier-Stokes equations with the top lid driven at a constant velocity 1 unit/s. An unstructured triangular mesh with 2211 nodes was employed for spatial discretization. From the full model simulation, 256 snapshots were obtained at spaced time intervals $\Delta t = 0.01$ for each of the $u$, $v$ and $p$ solution variables. We selected the first 160 snapshots for training, and predicted the solution of the problem at time $t=7.68$.

\begin{figure}[htbp!]
\centering
\begin{tabular}{cc}
\begin{minipage}{0.45\linewidth}
\includegraphics[width = \linewidth,angle=0,clip=true]{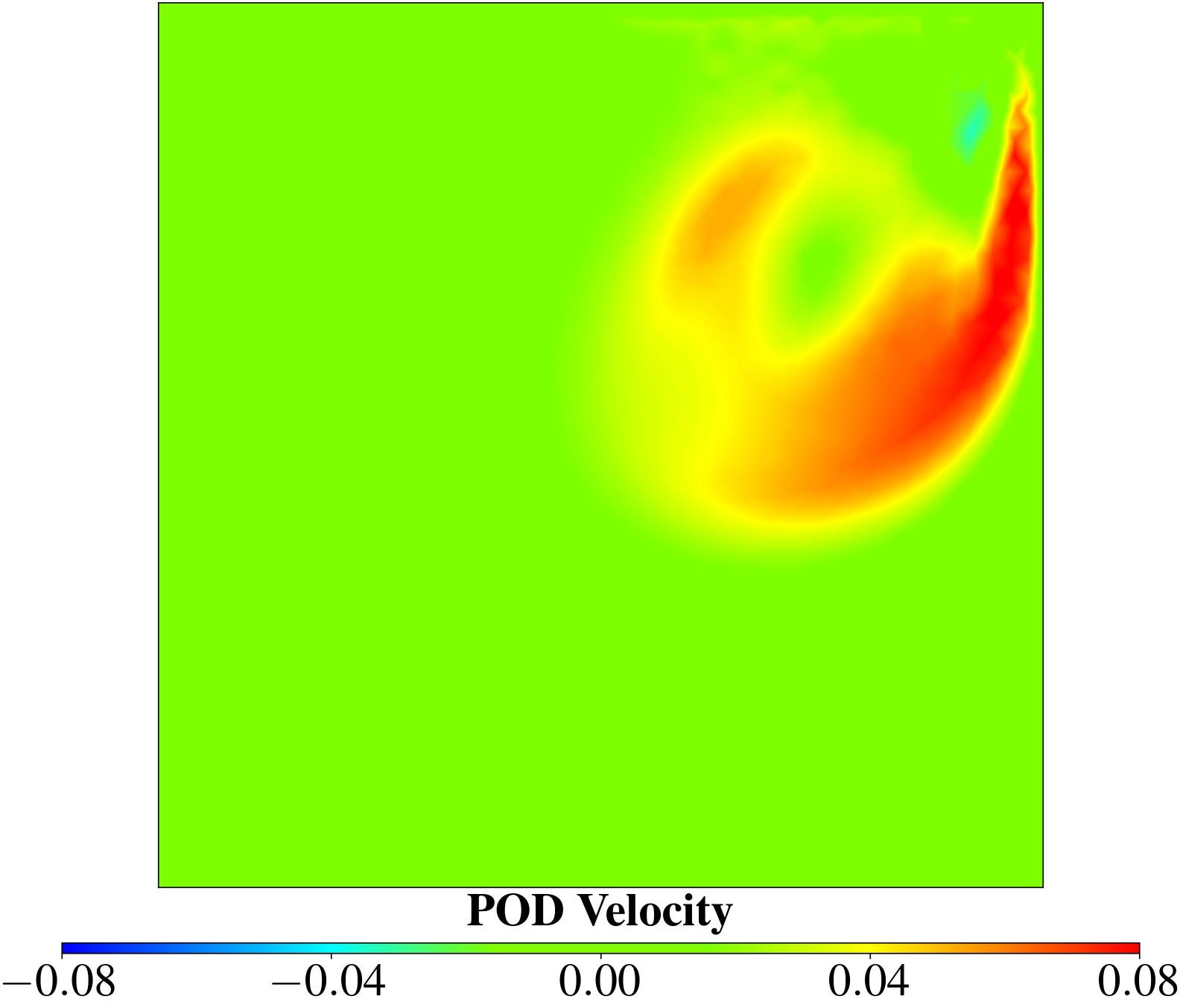}  
\end{minipage} 
&
\begin{minipage}{0.45\linewidth}
\includegraphics[width = \linewidth,angle=0,clip=true]{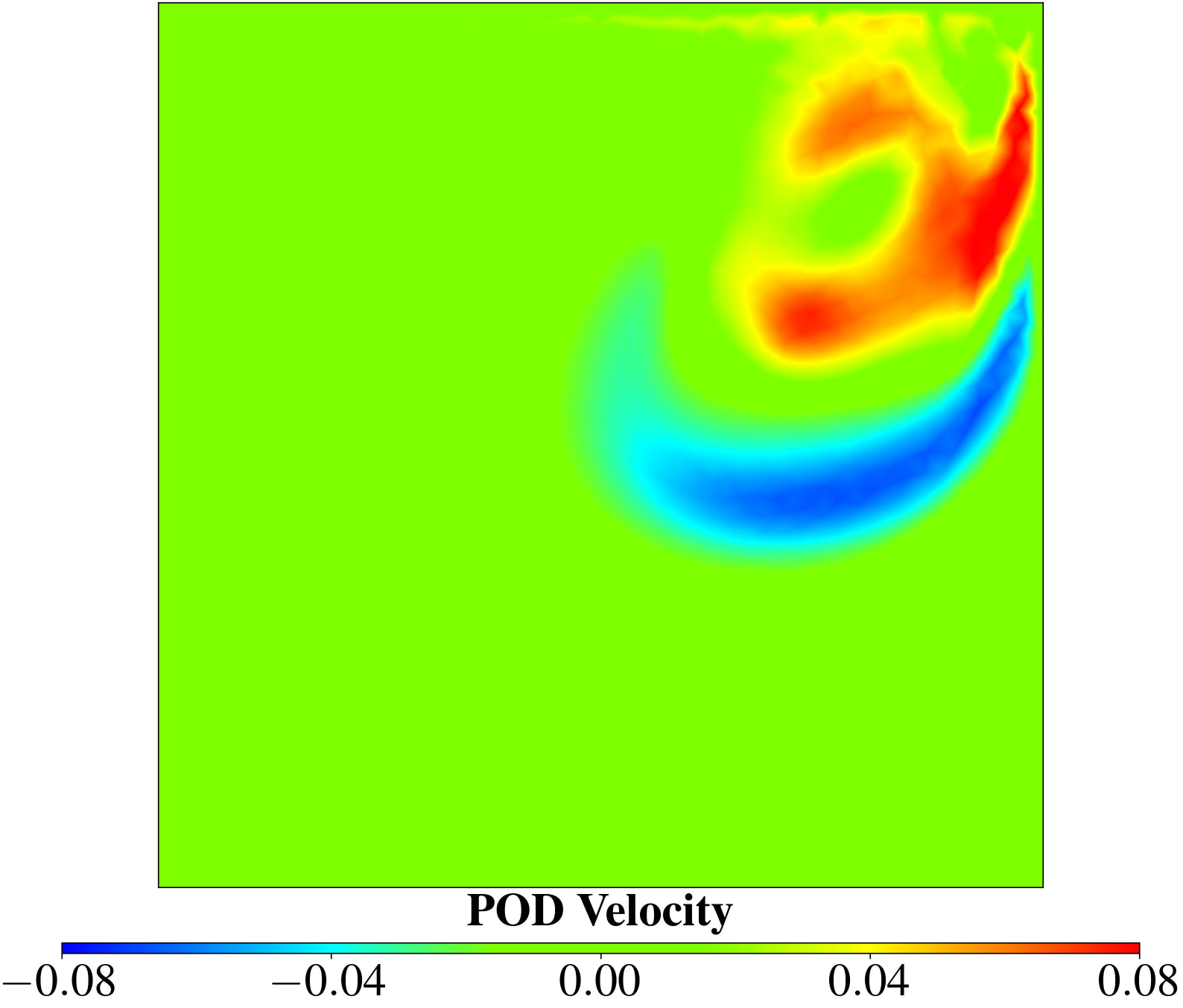}  
\end{minipage} 
 \\
(a) {\small 1st POD basis.}&
(b) {\small 2nd POD basis.}
\\

\begin{minipage}{0.45\linewidth}
\includegraphics[width = \linewidth,angle=0,clip=true]{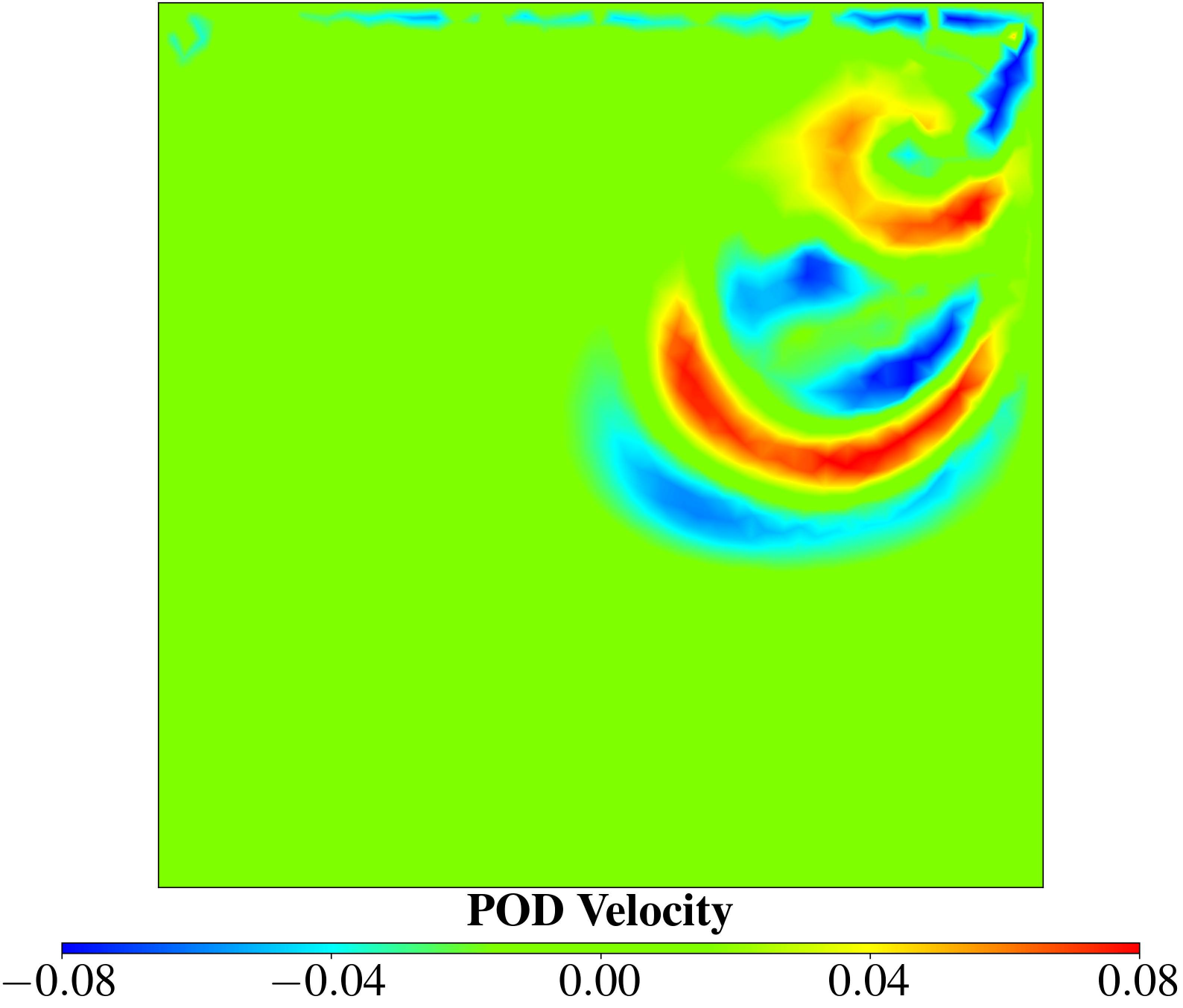}  
\end{minipage} 
&
\begin{minipage}{0.45\linewidth}
\includegraphics[width = \linewidth,angle=0,clip=true]{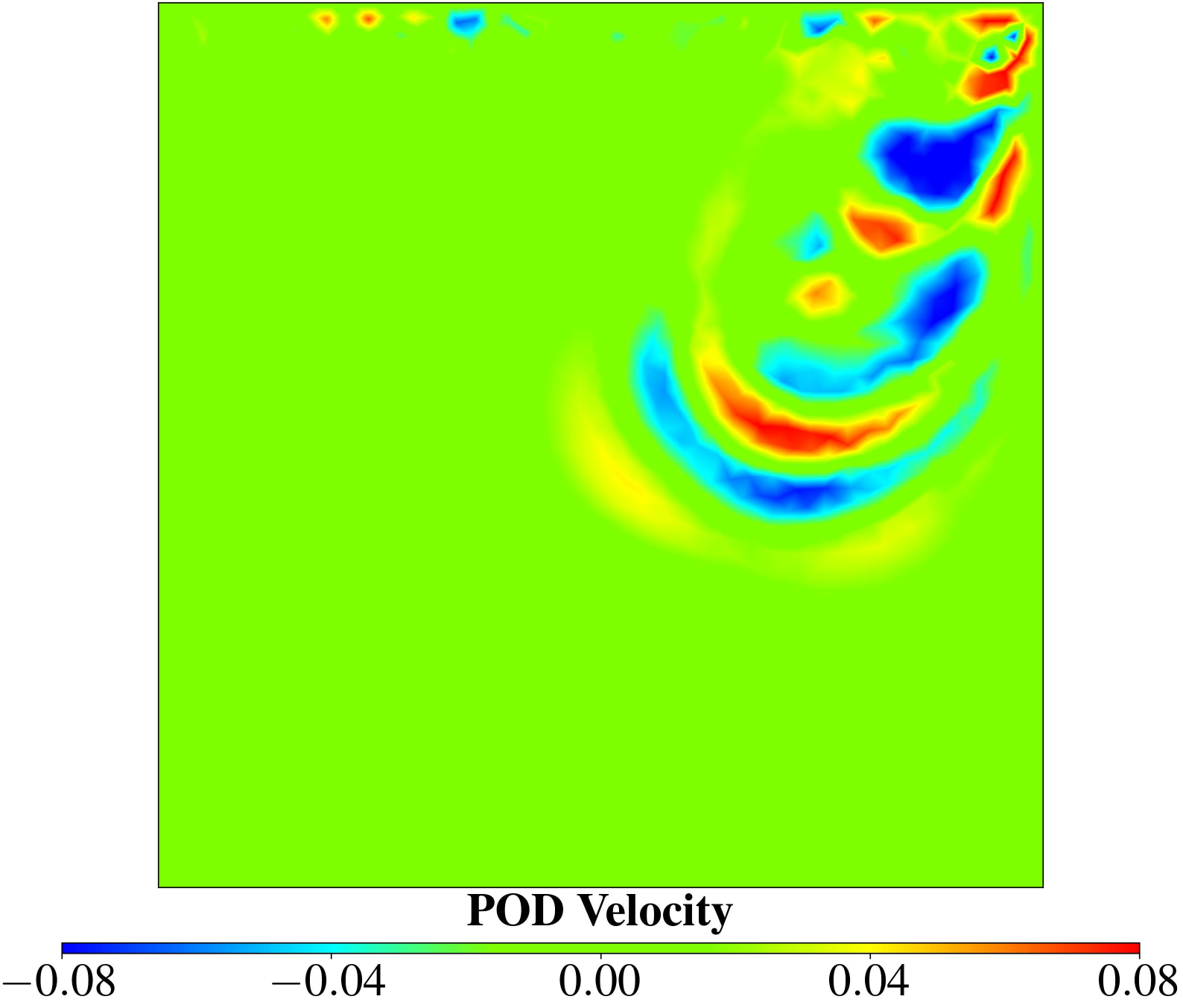}  
\end{minipage} 
 \\
(c) {\small 4st POD basis.}&
(d) {\small 8st POD basis.}
\\
\end{tabular}
\caption{\textbf{Lid-driven cavity flow at $\boldsymbol{Re=4000}$.} The figure shows some of the first 8 basis functions of the problem.}
\label{fig:cavity_basis}
\end{figure}

\begin{figure}[htbp!]
\centering
\begin{tabular}{cc}
\begin{minipage}{0.45\linewidth}
\includegraphics[width = \linewidth,angle=0,clip=true]{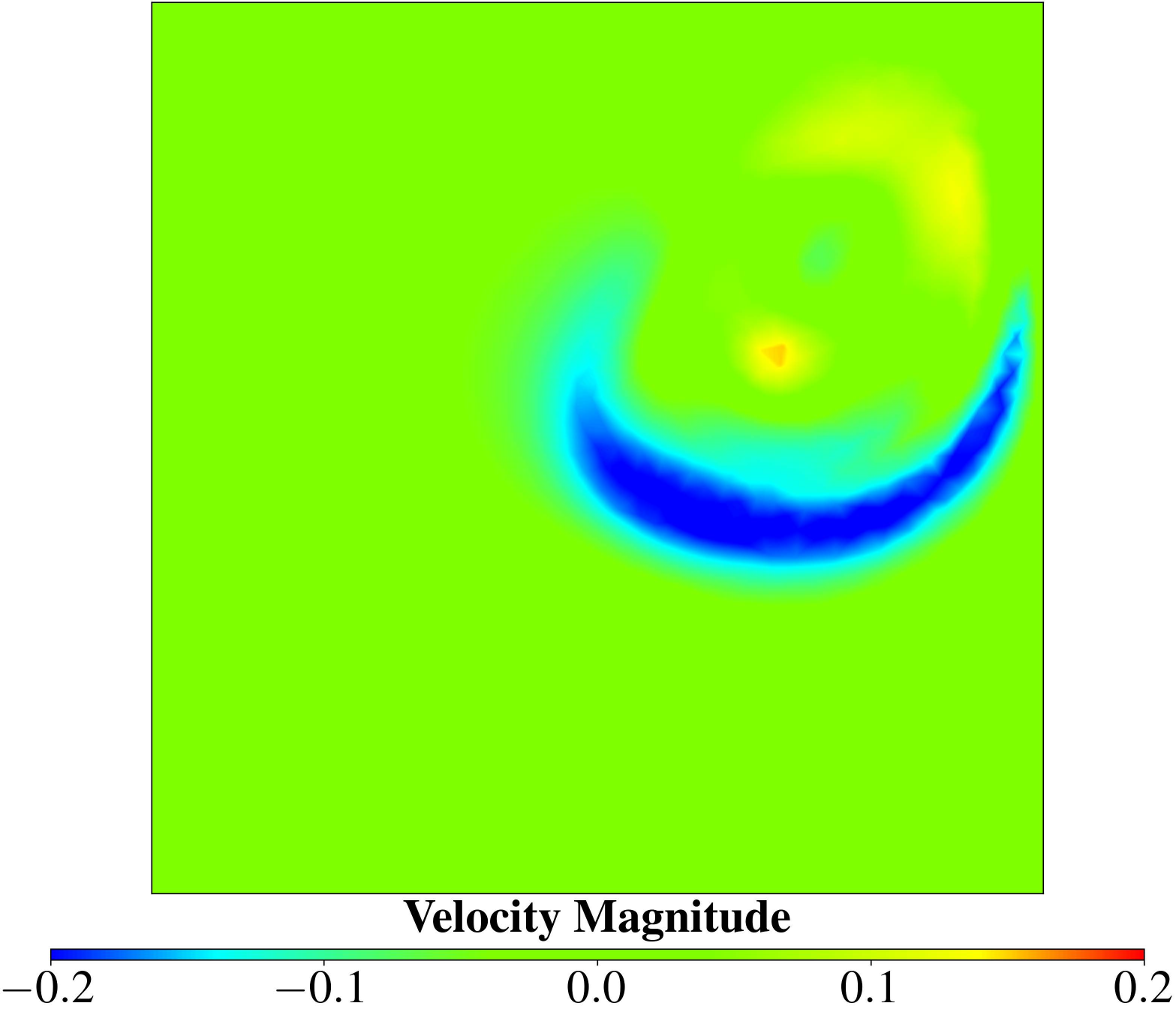}  
\end{minipage} 
&
\begin{minipage}{0.45\linewidth}
\includegraphics[width = \linewidth,angle=0,clip=true]{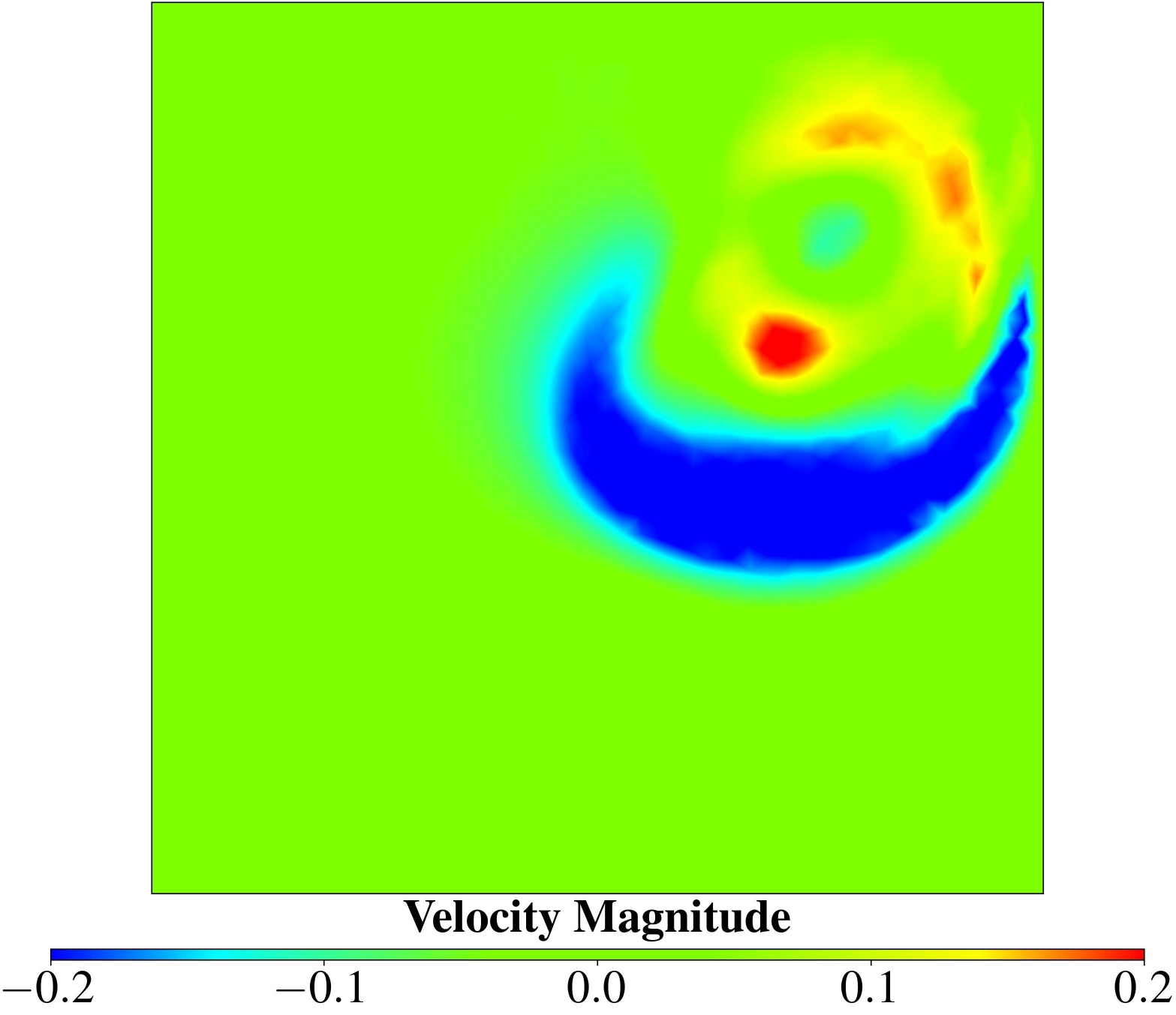}  
\end{minipage} 
 \\
(a) {\small Fully quantum framework, 1 POD basis}&
(b) {\small Classical framework, 1 POD basis}
\\

\begin{minipage}{0.45\linewidth}
\includegraphics[width = \linewidth,angle=0,clip=true]{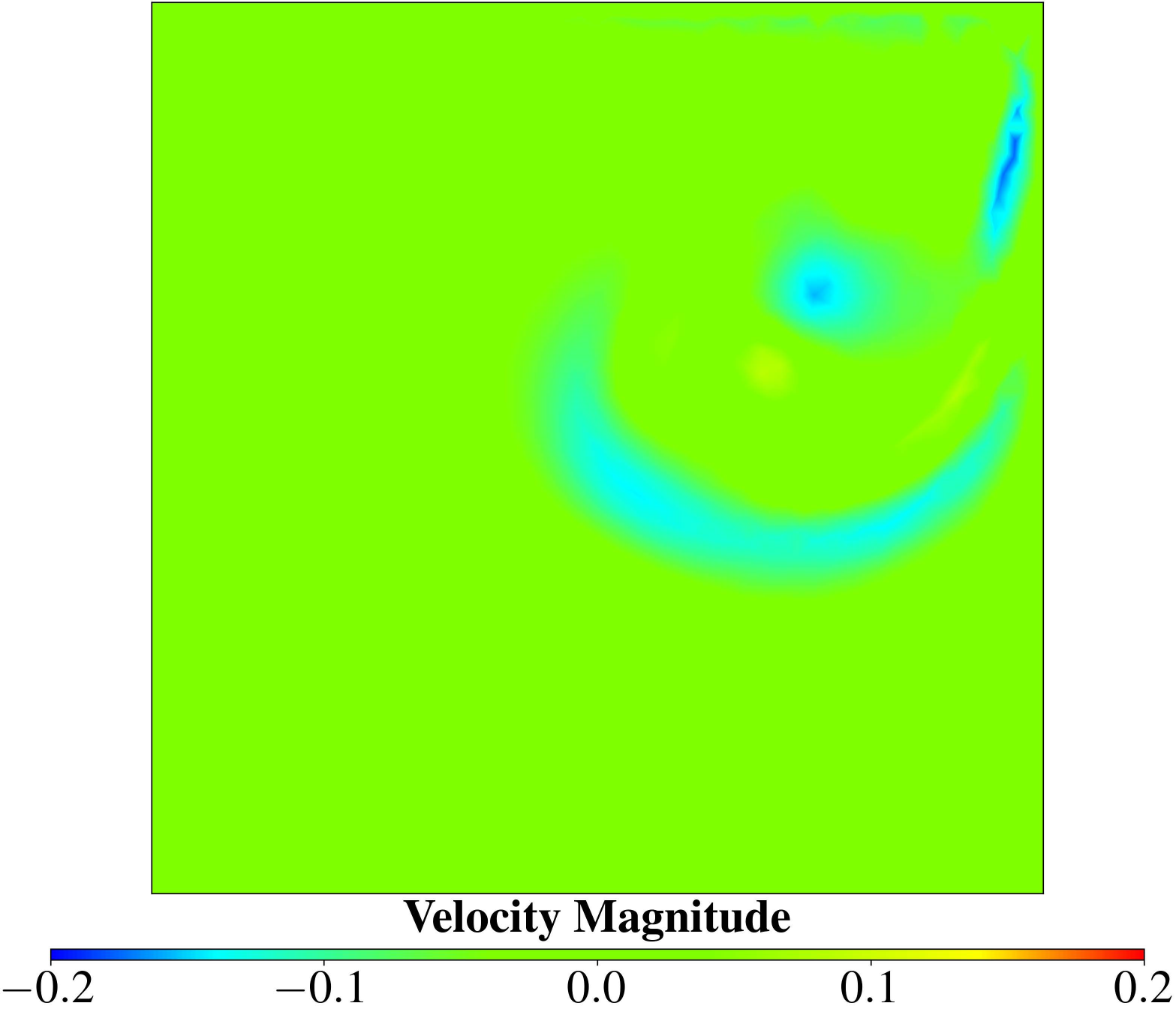}  
\end{minipage} 
&
\begin{minipage}{0.45\linewidth}
\includegraphics[width = \linewidth,angle=0,clip=true]{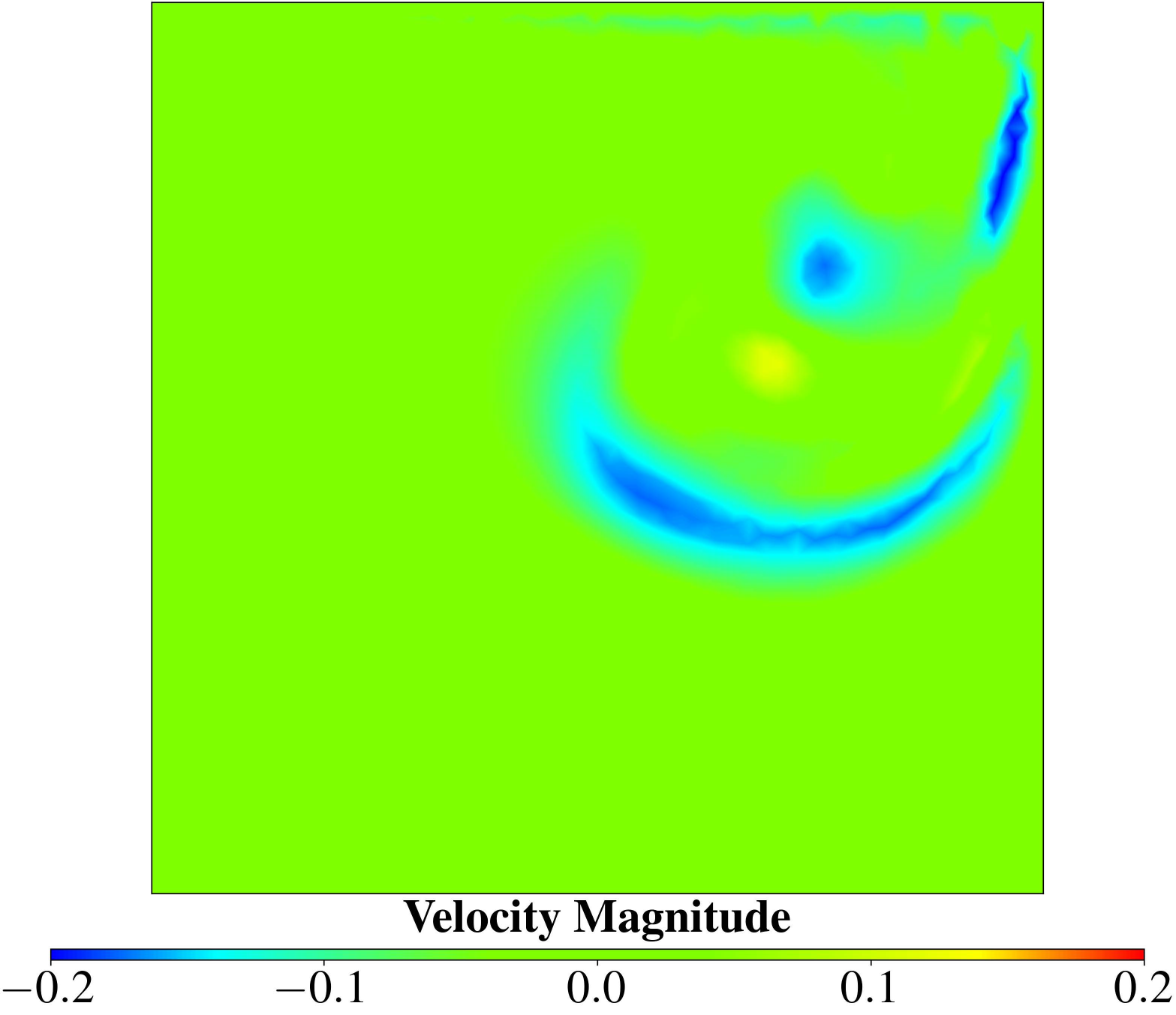}  
\end{minipage} 
 \\
(c) {\small Fully quantum framework, 2 POD bases}&
(d) {\small Classical framework, 2 POD bases}
\\
\begin{minipage}{0.45\linewidth}
\includegraphics[width = \linewidth,angle=0,clip=true]{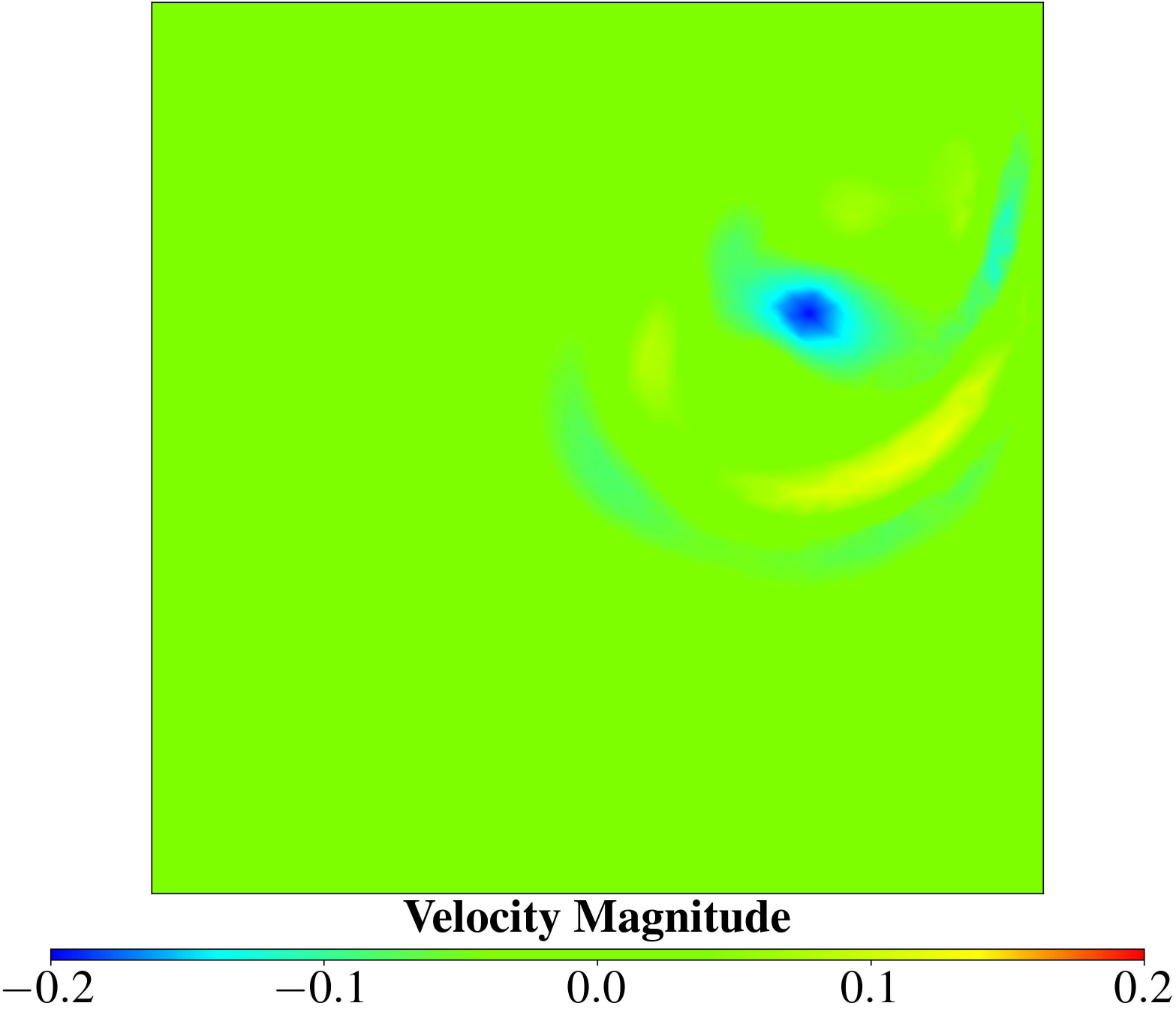}  
\end{minipage} 
&
\begin{minipage}{0.45\linewidth}
\includegraphics[width = \linewidth,angle=0,clip=true]{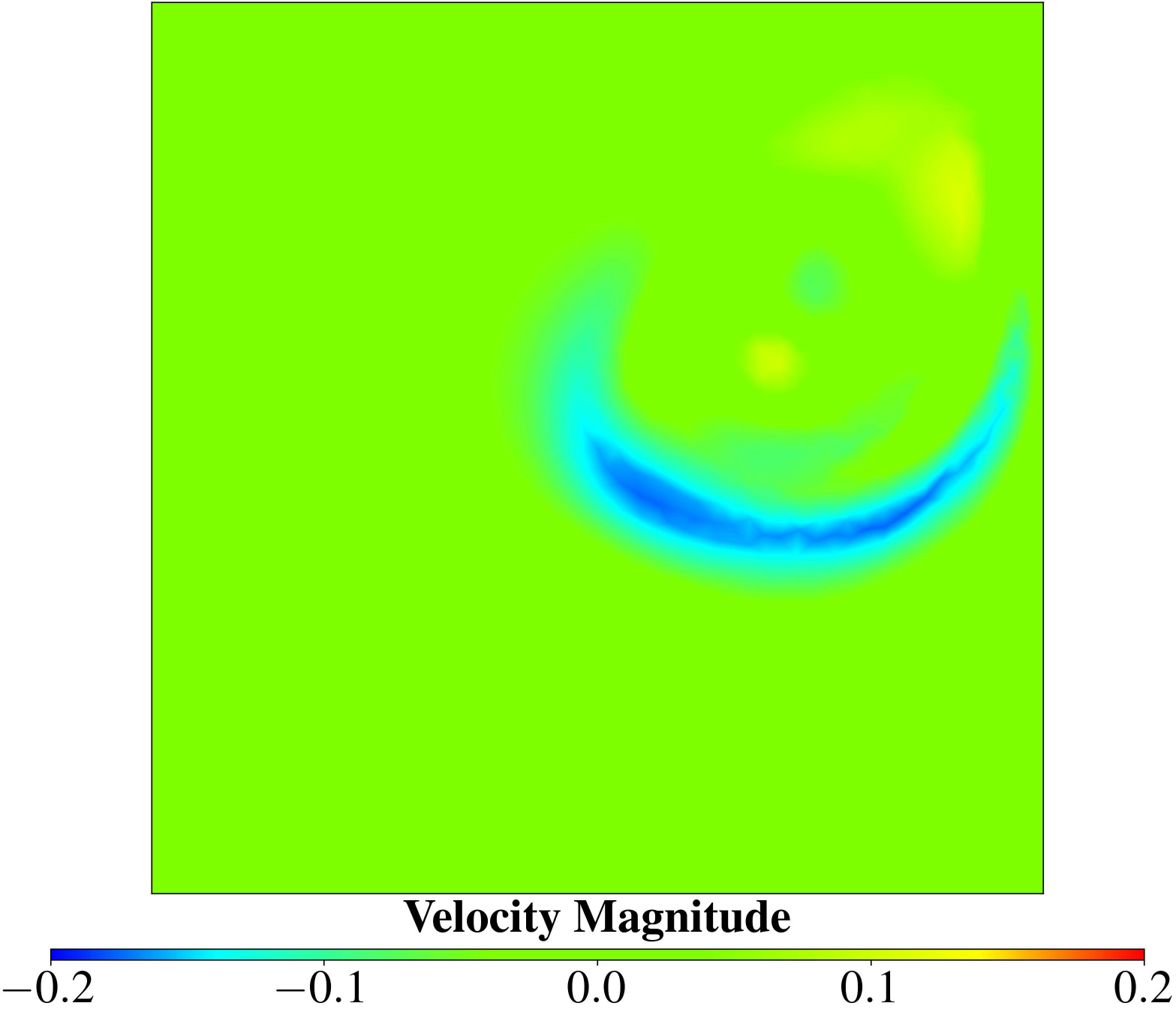}  
\end{minipage} 
 \\
(e) {\small Fully quantum framework, 3 POD bases}&
(f) {\small Classical framework, 3 POD bases}
\\
\end{tabular}
\caption{\textbf{Lid-driven cavity flow at $\boldsymbol{Re=4000}$}. The error of fully quantum framework and classical framework at time levels $t=7.68$ under the ROM rank 1, 2 and 3.}
\label{fig:1c_error}
\end{figure}

A subset of the first 40 POD basis functions is shown in Figure \ref{fig:cavity_basis}. Figure \ref{fig:1c_error} presents the errors of the simulated flow patterns using NIROMs (fully quantum POD-DKL framework and classical framework) at time instance $t=7.68$. The number of POD basis functions is set to 1, 2, and 3 respectively. It can be seen that the fully quantum POD-DKL framework exhibits the same level of errors as classical methods. However, the training epoches of classical method is 10 times more than that of quantum method.

In the second numerical example, the 2-D fluid dynamics problem of the flow past a cylinder is analyzed. The computational domain is configured as a 2-D rectangular region $\Omega=[0,2]\times[0,0.4]$ with cylinder diameter $D=0.12$ centered at $(0.2,0.2)$, governed by the incompressible Navier-Stokes equations under uniform inflow velocity 1 unit/s. The fluid enters the computational domain from the left boundary, flows past the cylinder and through the right boundary in the end. Dirichlet boundary conditions are applied to the cylinder surface while no slip and zero outward flow conditions are applied to the upper and lower boundaries of the domain. A hybrid unstructured mesh containing 2106 nodes was implemented. At Reynolds number $Re=4000$, totally 800 consecutive snapshots were collected from the full-order model simulation at temporal intervals $\Delta t=0.01$ for velocity components $u$, $v$ and pressure $p$. Leveraging quantum state superposition properties, the first 512 snapshots were utilized for reduced-order model training, with predictive validation conducted at final time $t=8$.

\begin{figure}[ht]
\centering
\includegraphics[width = 0.6\linewidth,angle=0,clip=true]{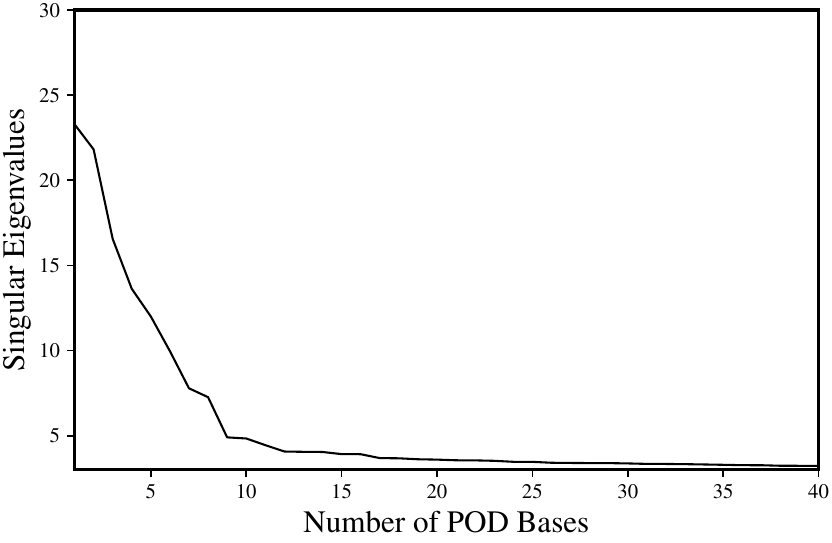}
\caption{\textbf{Flow past a cylinder at $\boldsymbol{Re=4000}$.} The graph shows the singular values of the 2-D flow past a cylinder problem.}\label{fig:sing_cylinder}
\end{figure}

\begin{figure}[htbp!]
\centering
\begin{tabular}{cc}
\begin{minipage}{0.45\linewidth}
\includegraphics[width = \linewidth,angle=0,clip=true]{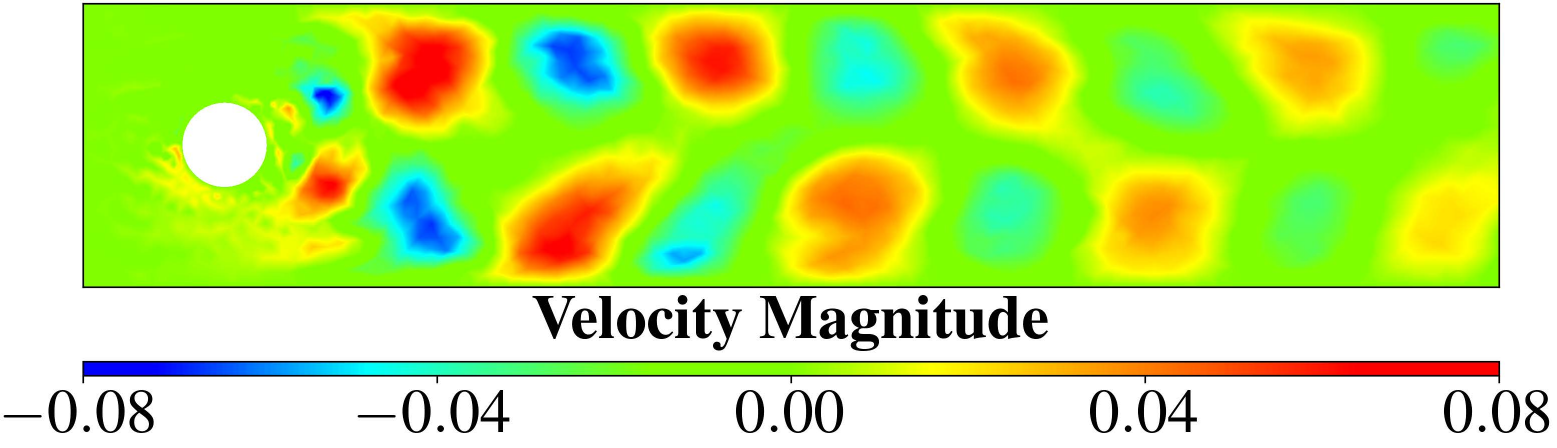}  
\end{minipage} 
&
\begin{minipage}{0.45\linewidth}
\includegraphics[width = \linewidth,angle=0,clip=true]{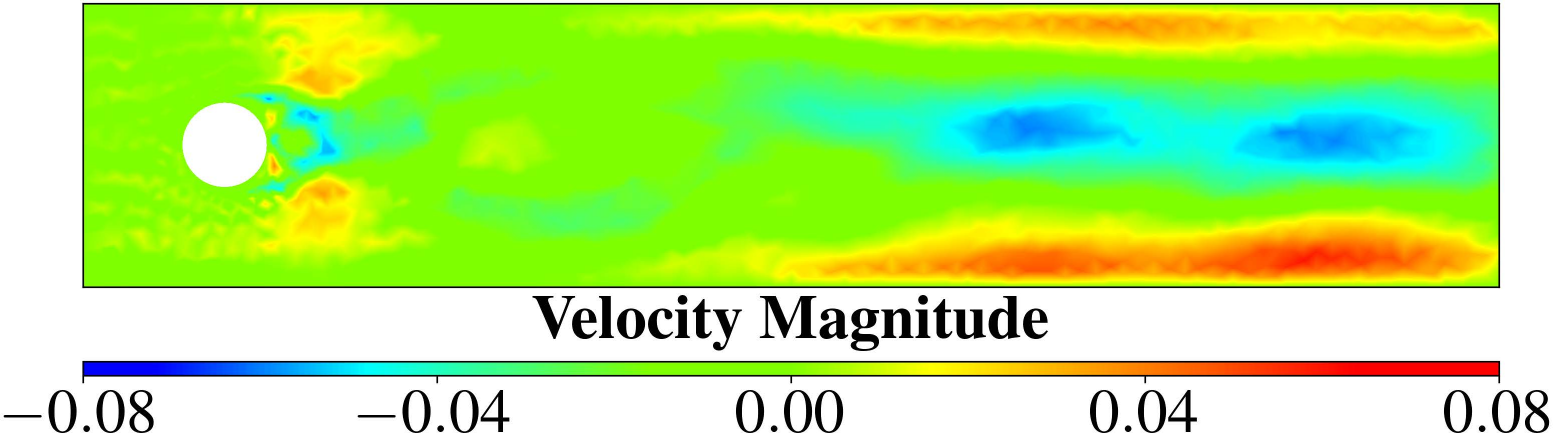}  
\end{minipage} 
 \\
(a) {\small 1st POD basis.}&
(b) {\small 2nd POD basis.}
\\

\begin{minipage}{0.45\linewidth}
\includegraphics[width = \linewidth,angle=0,clip=true]{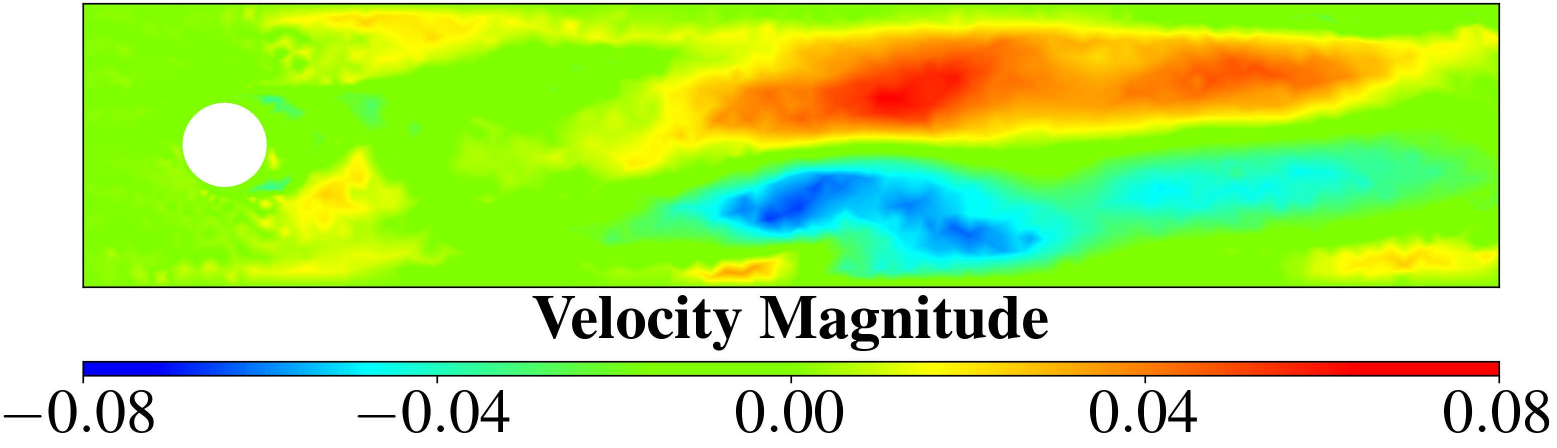}  
\end{minipage} 
&
\begin{minipage}{0.45\linewidth}
\includegraphics[width = \linewidth,angle=0,clip=true]{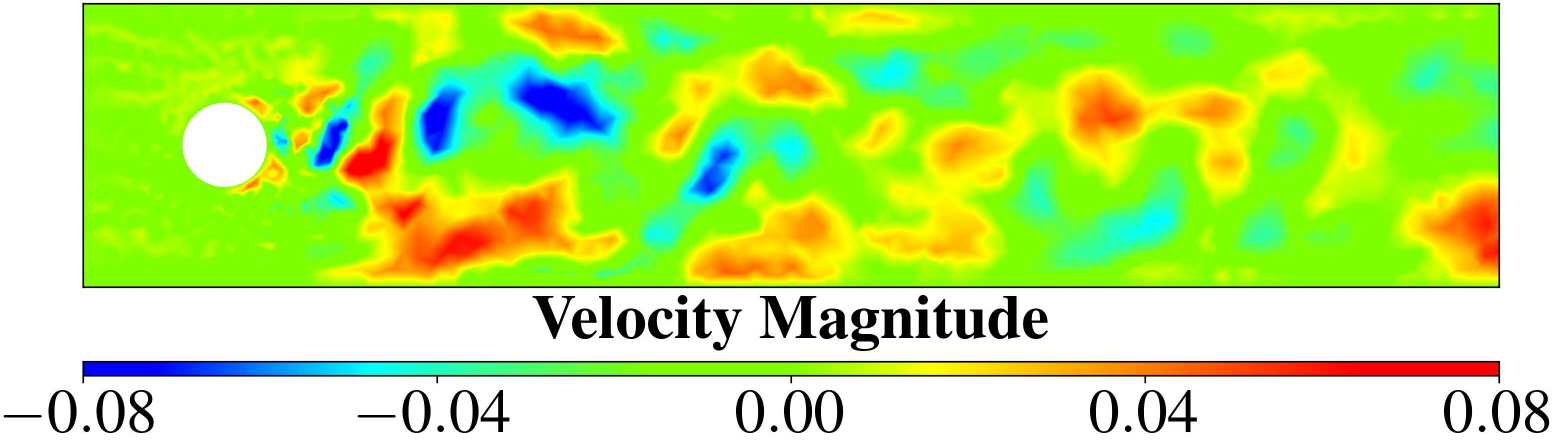}  
\end{minipage} 
 \\
(c) {\small 3rd POD basis.}&
(d) {\small 10st POD basis.}
\\
\begin{minipage}{0.45\linewidth}
\includegraphics[width = \linewidth,angle=0,clip=true]{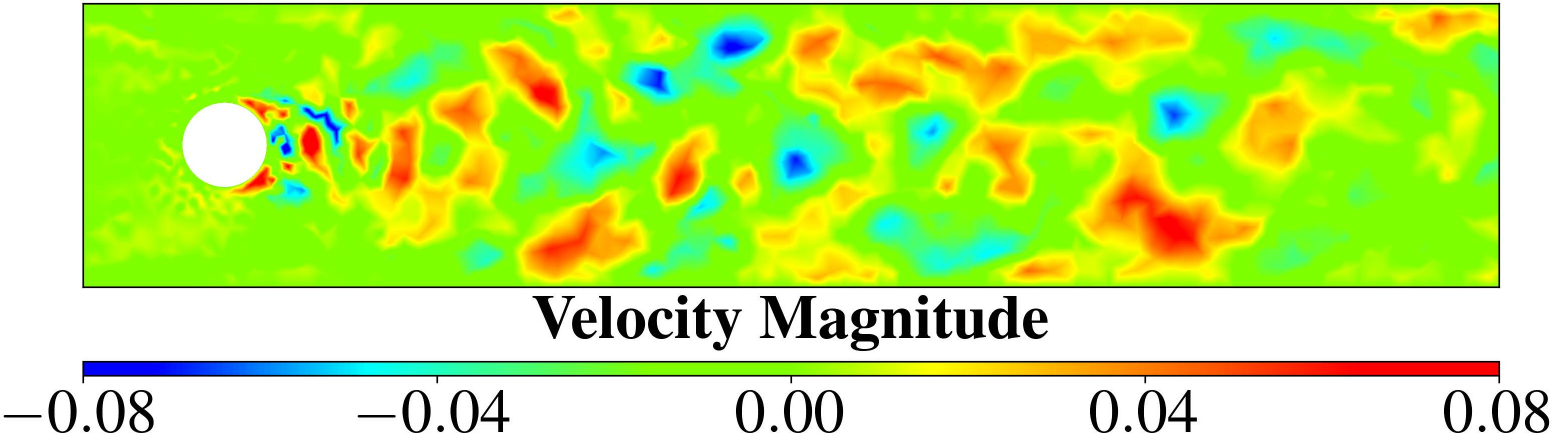}  
\end{minipage} 
&
\begin{minipage}{0.45\linewidth}
\includegraphics[width = \linewidth,angle=0,clip=true]{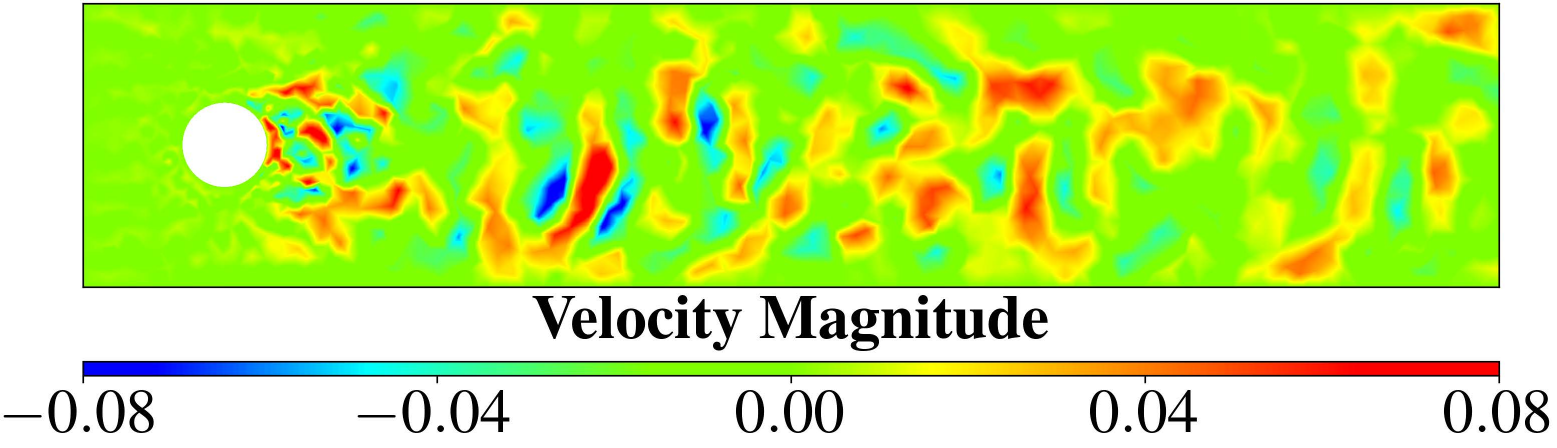}  
\end{minipage} 
 \\
(e) {\small 20st POD basis.}&
(f) {\small 40st POD basis.}
\\
\end{tabular}
\caption{\textbf{Flow past a cylinder.} The figure shows some of the first 40 bases functions of the problem.}
\label{fig:cylinder_basis}
\end{figure}

\begin{figure}[htbp!]
\centering
\begin{tabular}{cc}
\begin{minipage}{0.48\linewidth}
\includegraphics[width = \linewidth,angle=0,clip=true]{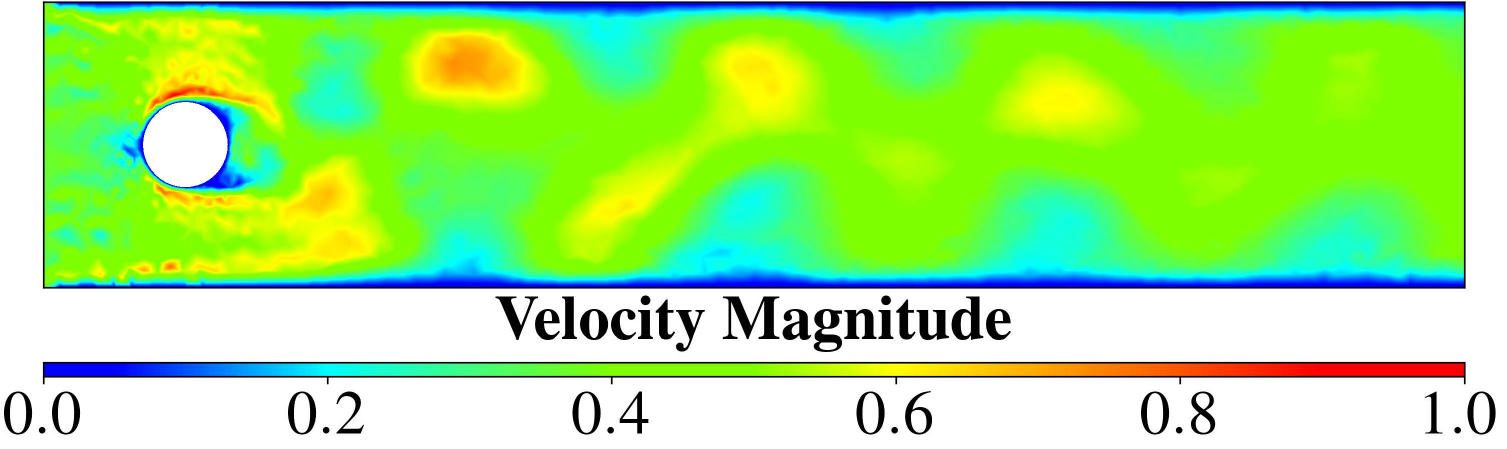}  
\end{minipage} 
&
\begin{minipage}{0.48\linewidth}
\includegraphics[width = \linewidth,angle=0,clip=true]{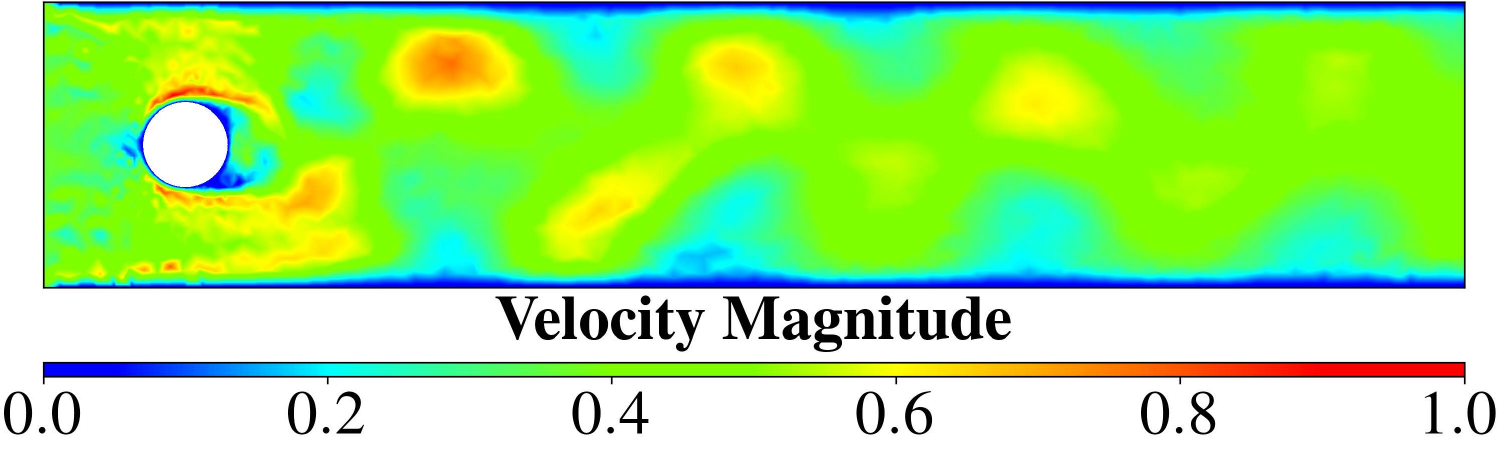}  
\end{minipage} 
 \\
(a) {\small Fully quantum framework, 2 POD bases}
&
(b) {\small Classical framework, 2 POD bases}
\\
\begin{minipage}{0.48\linewidth}
\includegraphics[width = \linewidth,angle=0,clip=true]{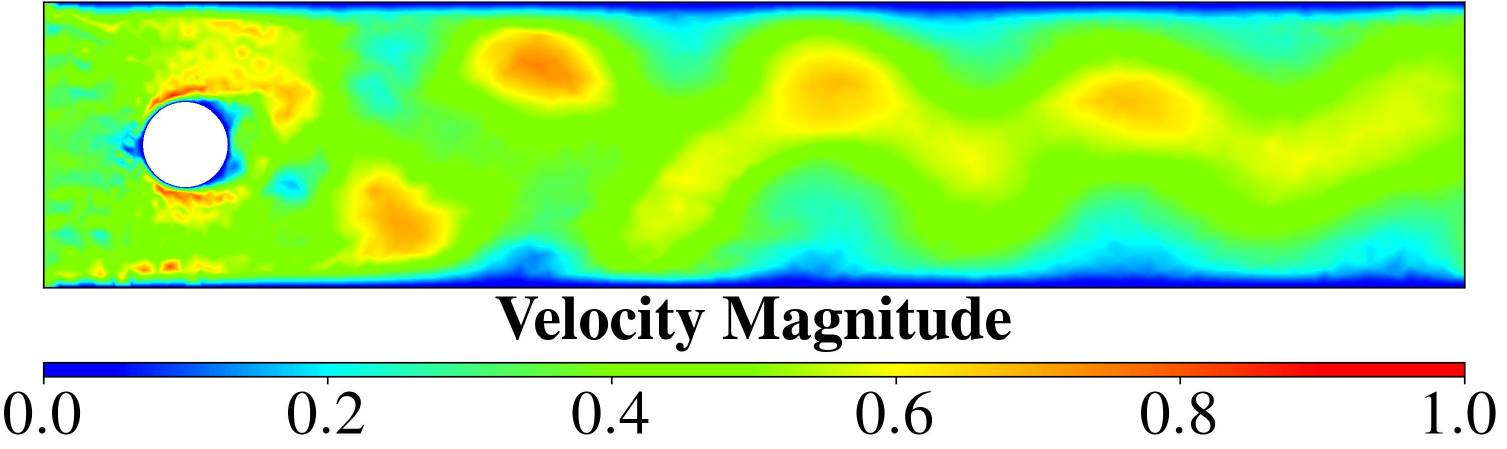}  
\end{minipage} 
&
\begin{minipage}{0.48\linewidth}
\includegraphics[width = \linewidth,angle=0,clip=true]{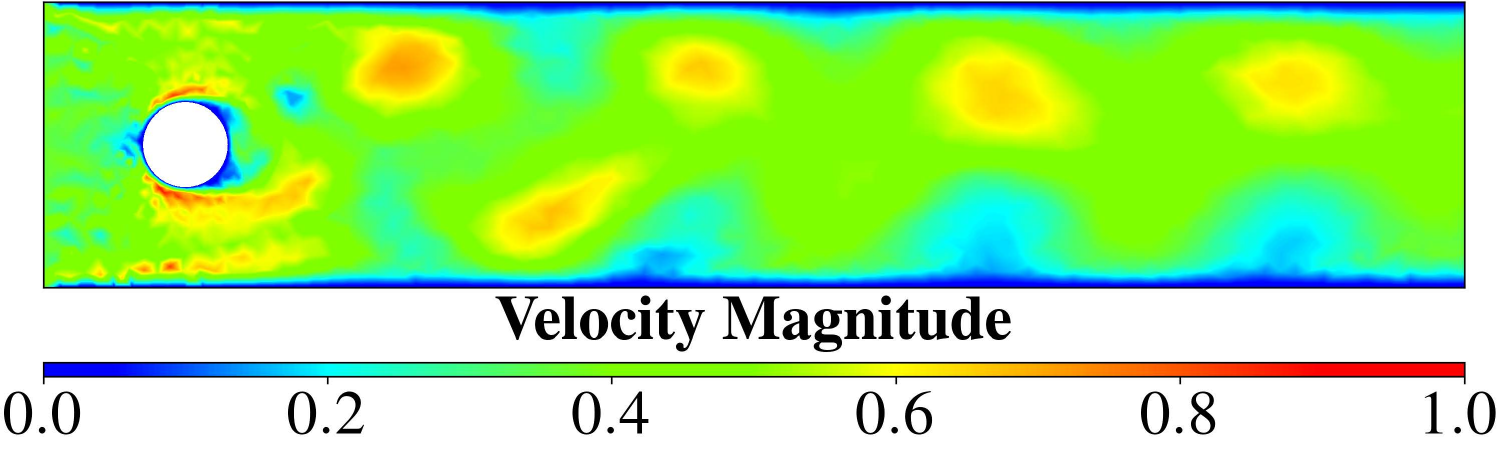}  
\end{minipage} 
 \\
(c) {\small Fully quantum framework, 5 POD bases}
&
(d) {\small Classical framework, 5 POD bases}

\\
\begin{minipage}{0.48\linewidth}
\includegraphics[width = \linewidth,angle=0,clip=true]{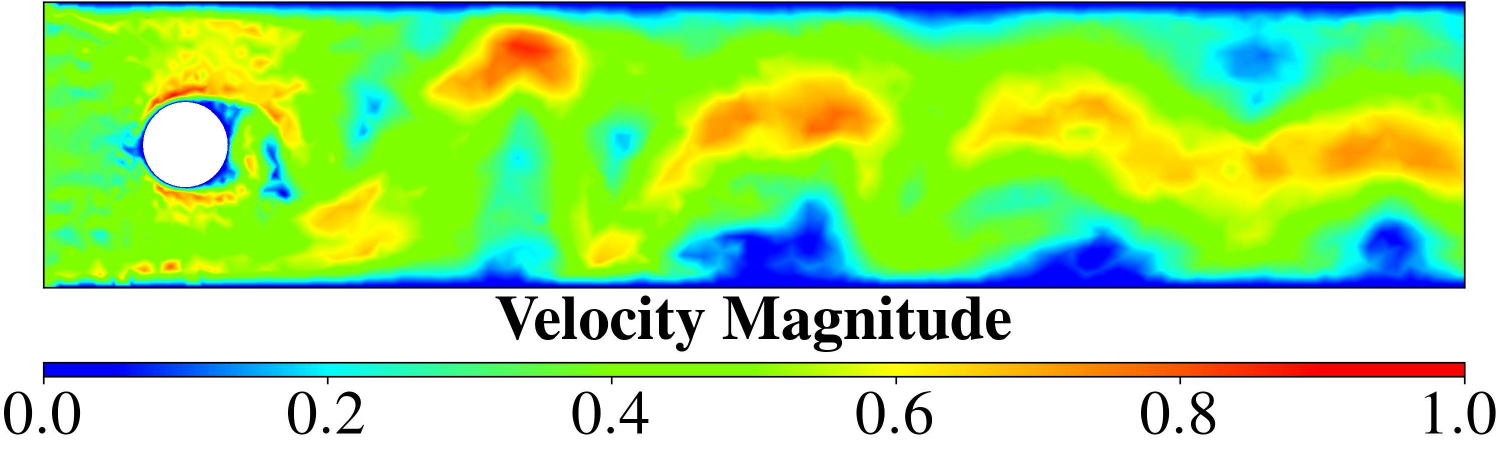}  
\end{minipage} 
&
\\
(e) {\small Full model solution}
&\,
\end{tabular}

\caption{\textbf{Flow past a cylinder at $\boldsymbol{Re=4000}$}. The solutions of fully quantum framework and classical framework at time levels $t=8$ using 2 and 5 POD bases.}
\label{fig:2c_solution}
\end{figure}

\begin{figure}[htbp!]
\centering
\begin{tabular}{cc}
\begin{minipage}{0.48\linewidth}
\includegraphics[width = \linewidth,angle=0,clip=true]{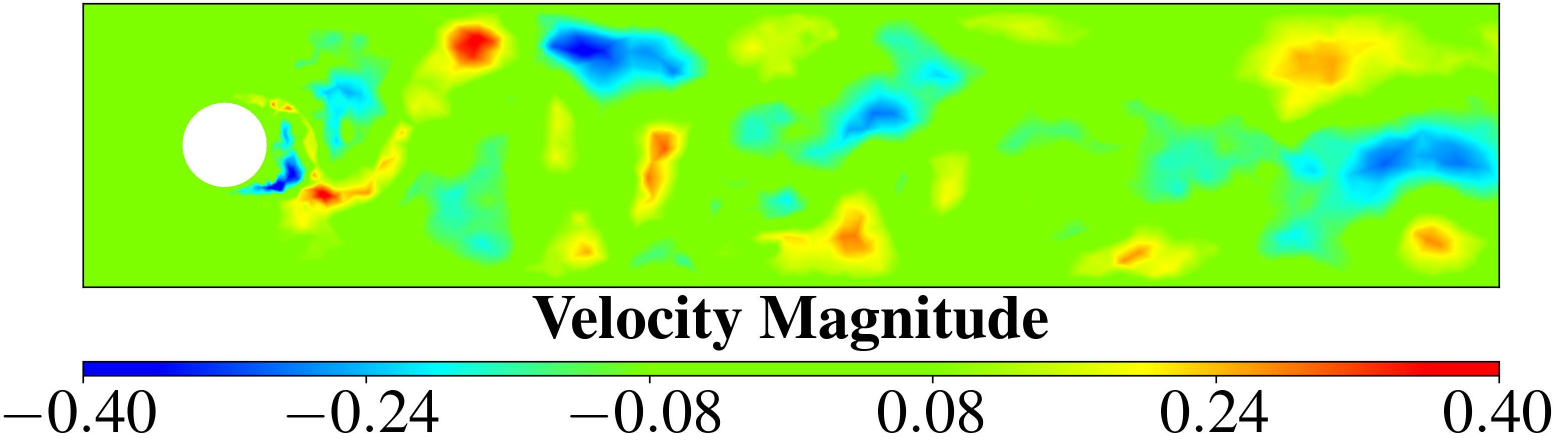}  
\end{minipage} 
&
\begin{minipage}{0.48\linewidth}
\includegraphics[width = \linewidth,angle=0,clip=true]{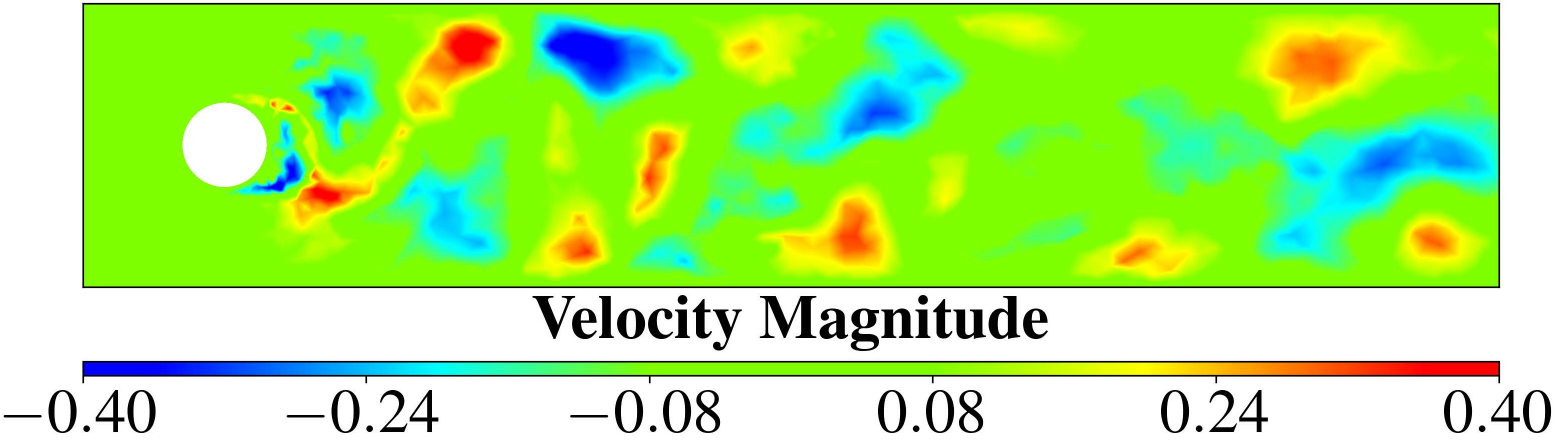}  
\end{minipage} 
 \\
(a) {\small Fully quantum framework, 2 POD bases}
&
(b) {\small Classical framework, 2 POD bases}
\\
\begin{minipage}{0.48\linewidth}
\includegraphics[width = \linewidth,angle=0,clip=true]{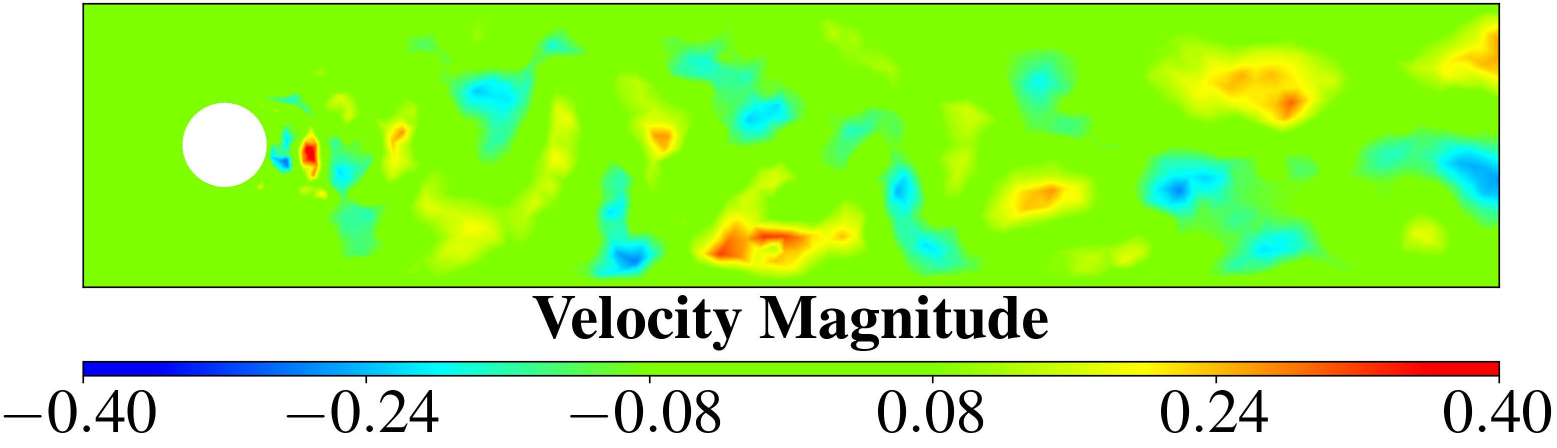}  
\end{minipage} 
&
\begin{minipage}{0.48\linewidth}
\includegraphics[width = \linewidth,angle=0,clip=true]{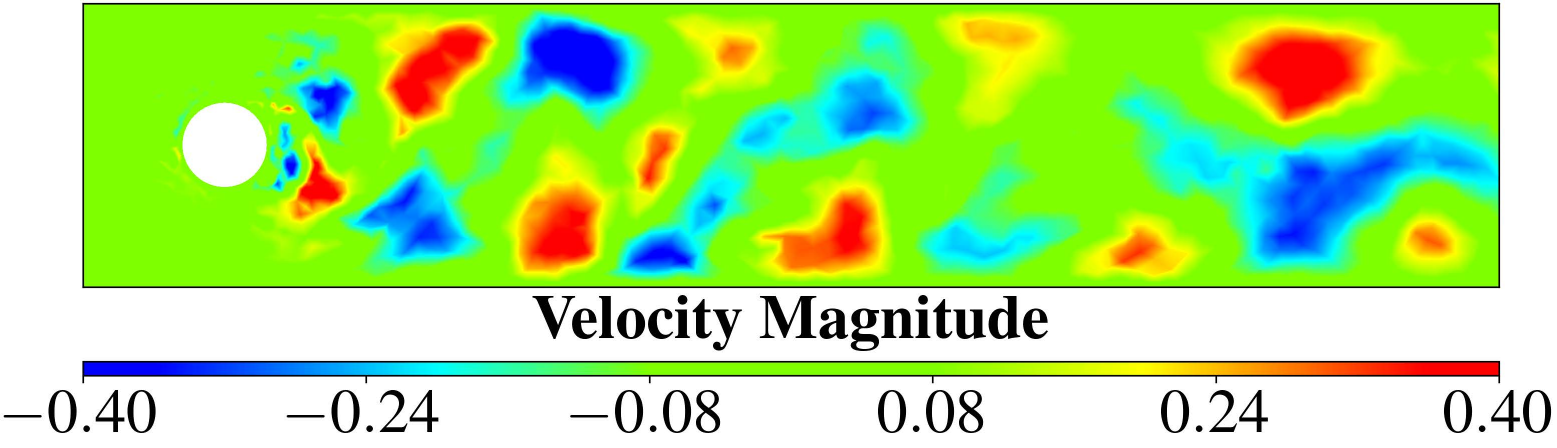}  
\end{minipage} 
 \\
(c) {\small Fully quantum framework, 5 POD bases}
&
(d) {\small Classical framework, 5 POD bases}
\end{tabular}

\caption{\textbf{Flow past a cylinder at $\boldsymbol{Re=4000}$}. The error of fully quantum framework and classical framework at time levels $t=5$ using 2 and 5 bases.}
\label{fig:2c_error}
\end{figure}

\begin{figure}[htbp!]
\centering
\begin{tabular}{c}
\includegraphics[width = 0.6\linewidth,angle=0,clip=true]{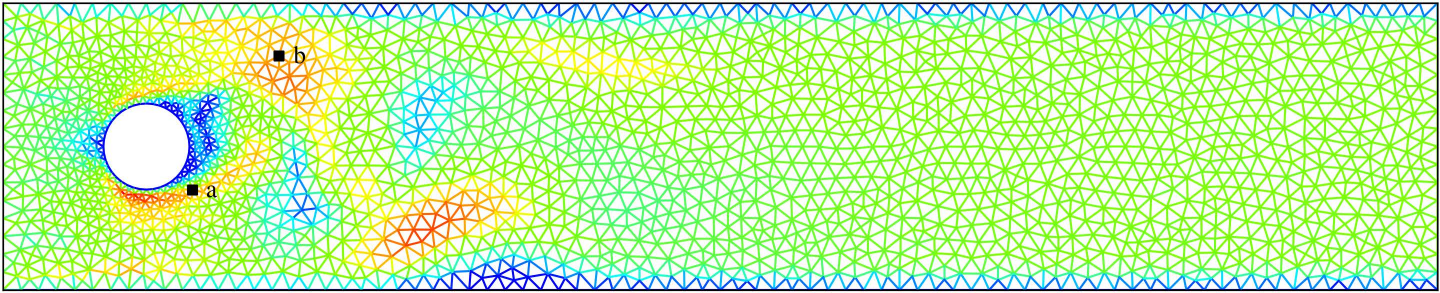}  
 \\
(a) {\small Locations a (0.264, 0.140) and b (0.385, 0.327).}
\\
\includegraphics[width = 0.7\linewidth,angle=0,clip=true]{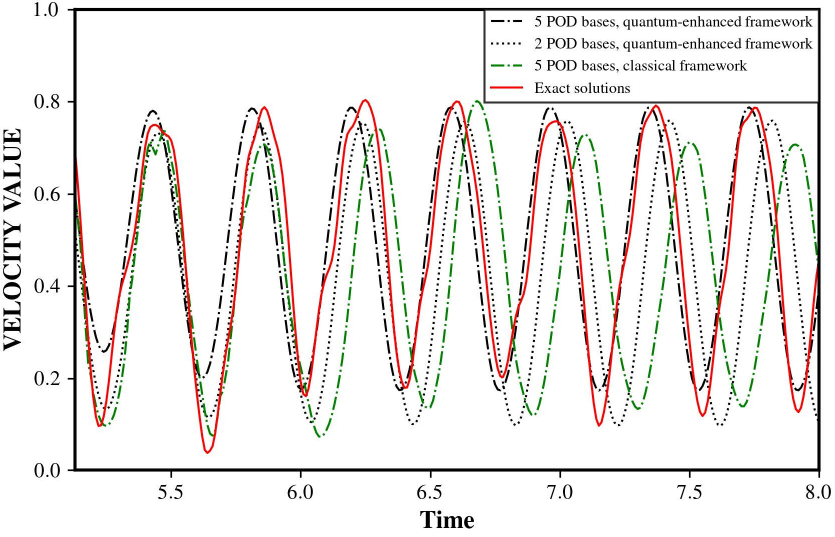}  
 \\
(b) {\small Fluid speed at a (0.264, 0.140).}
\\
\includegraphics[width = 0.7\linewidth,angle=0,clip=true]{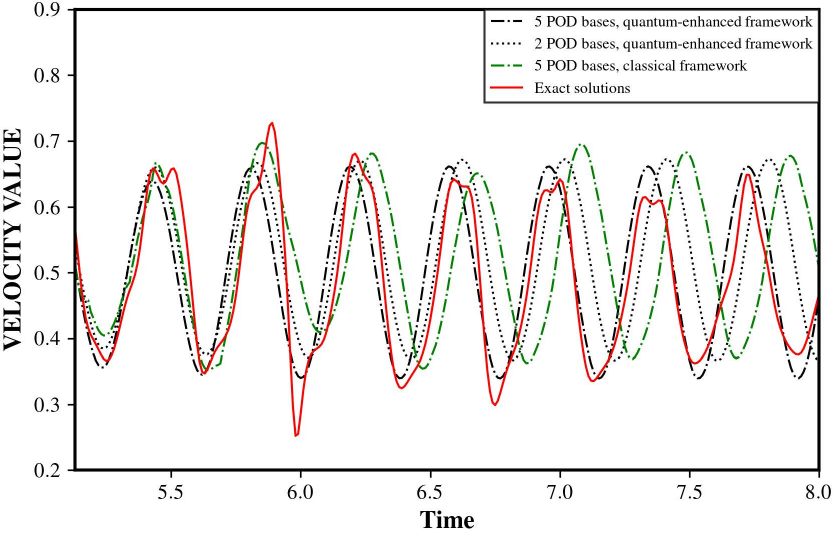}
\\
(c) {\small Fluid speed at b (0.385, 0.327).}
\end{tabular}

\caption{\textbf{Flow past a cylinder at $\boldsymbol{Re=4000}$}. The graphs show the flow speed predicted by the full model, the quantum-enhanced framework and the classical ROM framework at positions a (0.264, 0.140) and b (0.385, 0.327). These results were obtained using a reduced order model with 2 and 5 POD functions. }
\label{fig: point_velocity}
\end{figure}

Figure \ref{fig:sing_cylinder} shows the singular values calculated with quantum method. The rapid decay of the first 15 singular values reveals that the dominant basis functions associated with them encode the primary energy features of the original dynamical system. A subset of the first 40 POD basis functions calculated with quantum method is shown in Figure \ref{fig:cylinder_basis}, where low-order functions describe macroscopic velocity fields, while higher-order ones resolve fine-scale flow structures. Figure \ref{fig:2c_solution} presents the simulated flow patterns at time instance $t=8$, with the NIROMs (fully quantum POD-DKL framework and classical framework) compared against the high-fidelity model. The number of POD basis functions is set to 2 and 5 respectively. The fully quantum framework performs well, with the velocity magnitude highly consistent with that of the high-fidelity model. It can be seen that the fully quantum POD-DKL framework accurately captures the vortex shedding frequency and wake structures. Figure \ref{fig:2c_error} demonstrate that the fully quantum POD-DKL framework maintains the lowest prediction error. Figure \ref{fig: point_velocity} shows the flow speed at two points in the domain using 2 POD bases and 5 POD bases. It is again shown that the better accuracy of velocity solution predicted by fully quantum framework.

In the third case, the NIROM is applied to simulate the lock-exchange problem with Reynolds number $Re = 4000$, which involves two fluids of different temperatures and densities initially separated by a lock gate. Upon removal of the lock gate, two opposing currents form and propagate horizontally along the tank. This laboratory-scale setup simulates the dynamics of gravity currents across multiple scales. The computational domain is a dimensionless rectangle $\Omega=[0,0.8]\times[0,0.1]$. Initial conditions are defined as velocity $u_0=0$ and pressure $p_0=0$. The isotropic dynamic viscosity is set to $1\times10^{-8}$. The computational mesh comprises 3250 nodes, and 256 snapshots are uniformly sampled from high-fidelity simulations over the time interval $[0,18.36]$. For this test case, the temperature serves as the varied physical parameter.

\begin{figure}[ht]
\centering
\includegraphics[width = 0.6\linewidth,angle=0,clip=true]{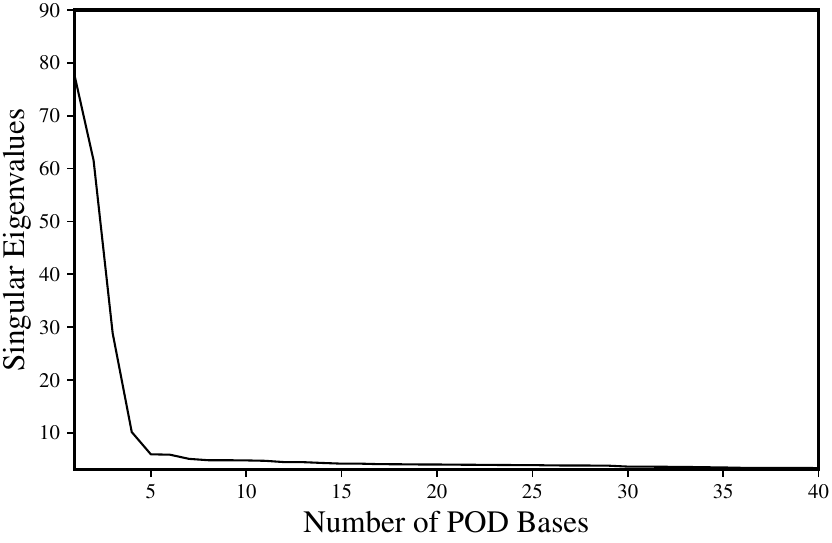}
\caption{\textbf{Lock exchange at $\boldsymbol{Re=4000}$.} The graph shows the singular values of the 2-D lock exchange problem.}\label{fig:sing}
\end{figure}

\begin{figure}[htbp!]
\centering
\begin{tabular}{cc}
\begin{minipage}{0.45\linewidth}
\includegraphics[width = \linewidth,angle=0,clip=true]{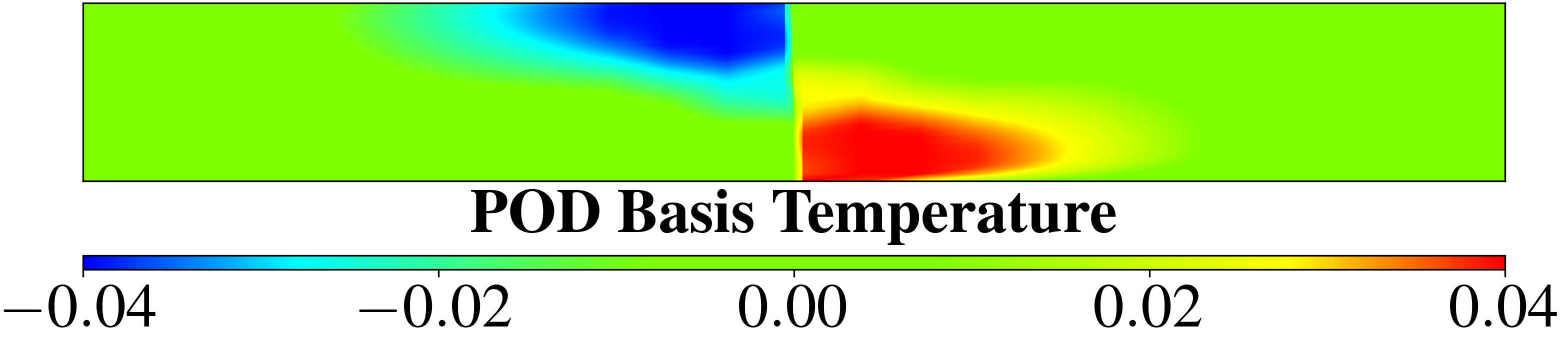}  
\end{minipage} 
&
\begin{minipage}{0.45\linewidth}
\includegraphics[width = \linewidth,angle=0,clip=true]{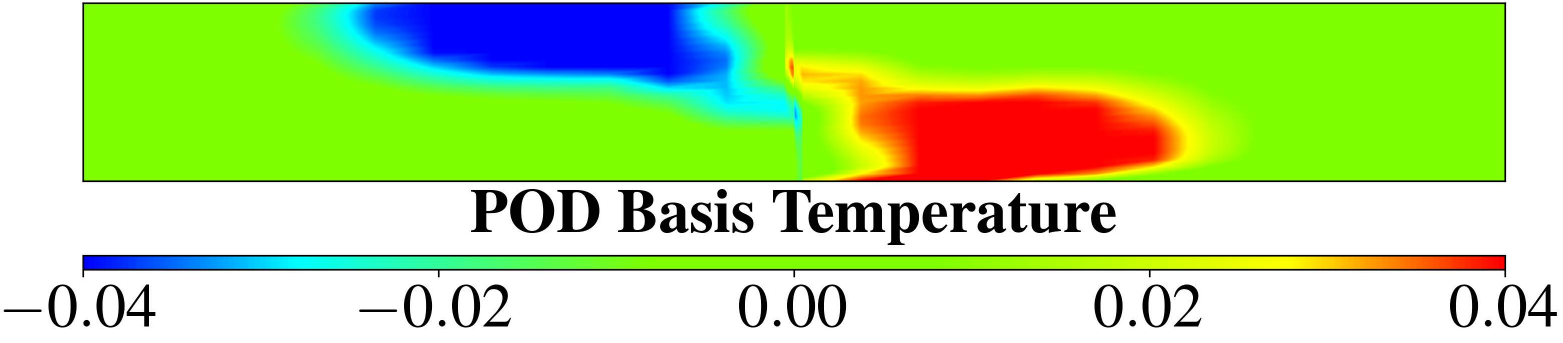}  
\end{minipage} 
 \\
(a) {\small 1st POD basis.}&
(b) {\small 2nd POD basis.}
\\

\begin{minipage}{0.45\linewidth}
\includegraphics[width = \linewidth,angle=0,clip=true]{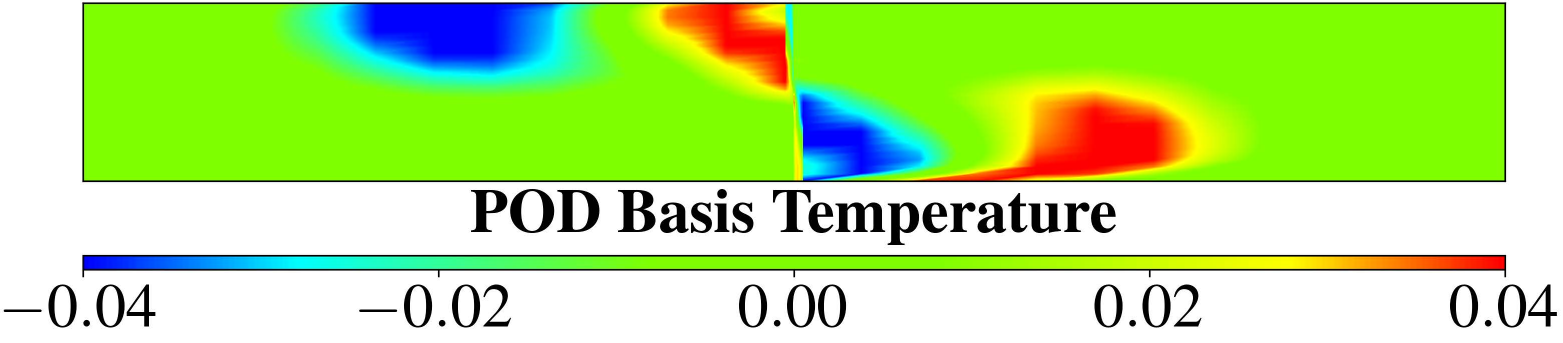}  
\end{minipage} 
&
\begin{minipage}{0.45\linewidth}
\includegraphics[width = \linewidth,angle=0,clip=true]{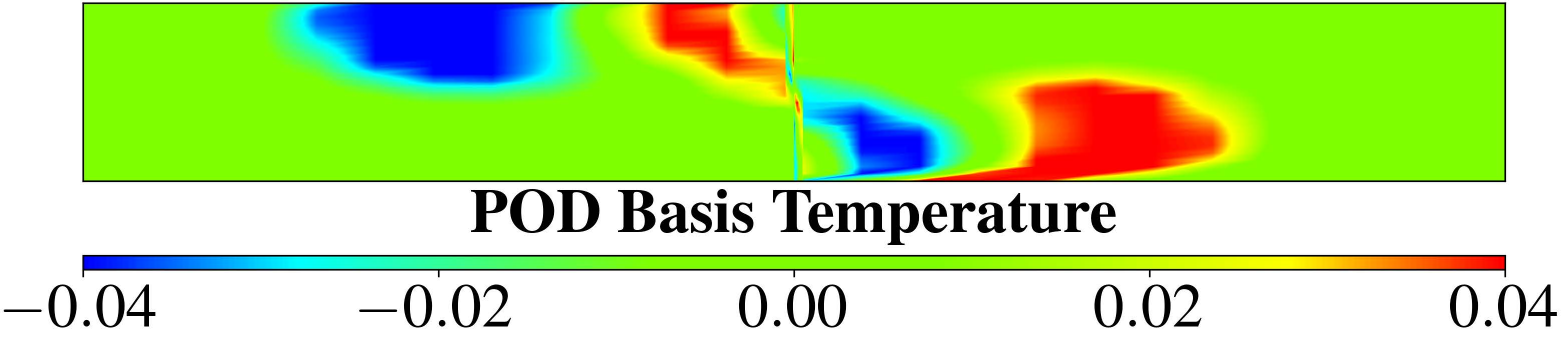}  
\end{minipage} 
 \\
(c) {\small 3rd POD basis.}&
(d) {\small 4st POD basis.}
\\
\begin{minipage}{0.45\linewidth}
\includegraphics[width = \linewidth,angle=0,clip=true]{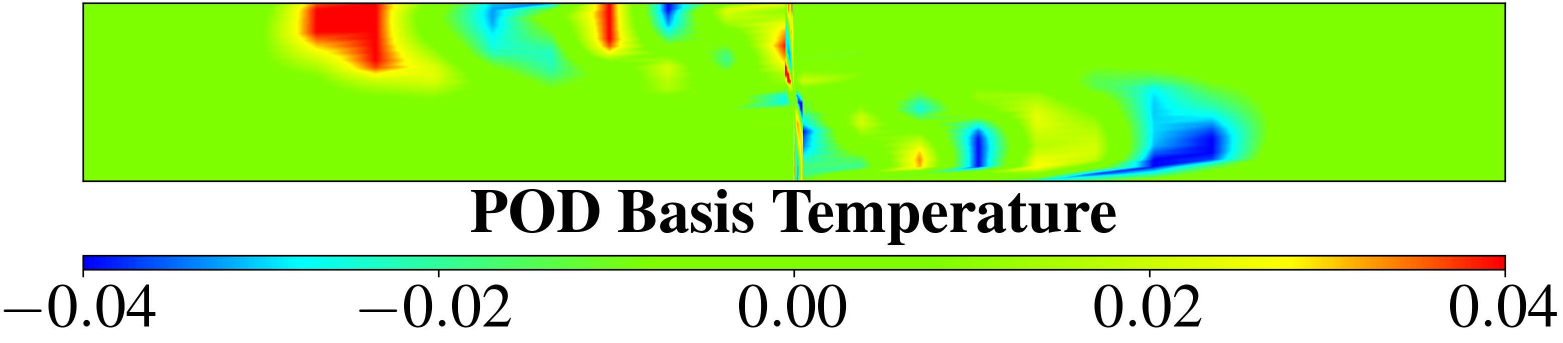}  
\end{minipage} 
&
\begin{minipage}{0.45\linewidth}
\includegraphics[width = \linewidth,angle=0,clip=true]{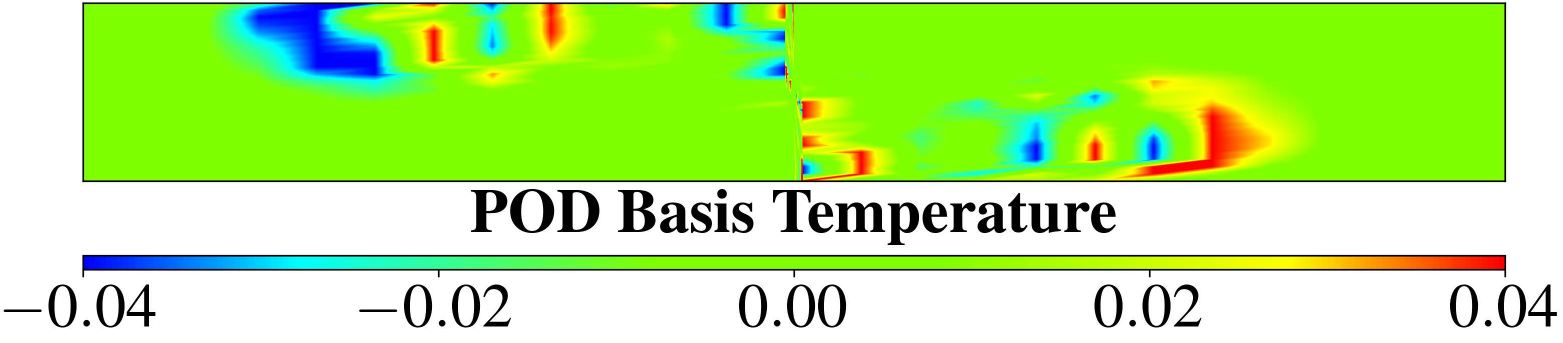}  
\end{minipage} 
 \\
(e) {\small 5st POD basis.}&
(f) {\small 15st POD basis.}
\\

\begin{minipage}{0.45\linewidth}
\includegraphics[width = \linewidth,angle=0,clip=true]{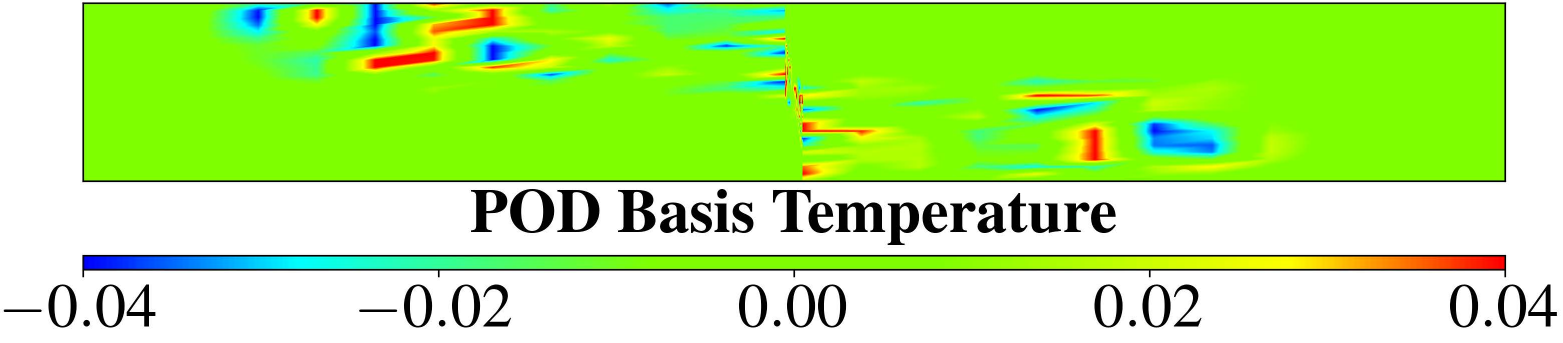}  
\end{minipage} 
&
\begin{minipage}{0.45\linewidth}
\includegraphics[width = \linewidth,angle=0,clip=true]{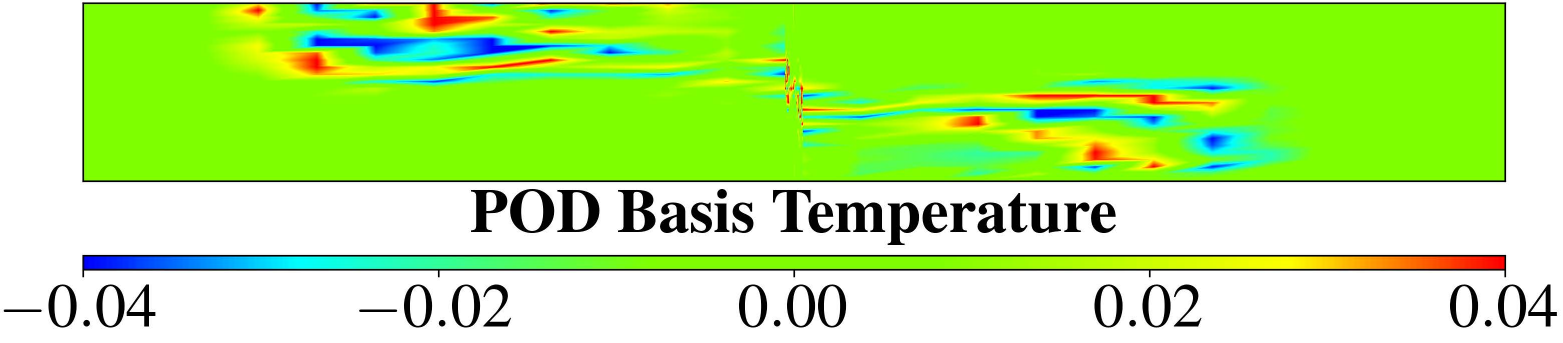}  
\end{minipage} 
 \\
(g) {\small 30th POD basis.}&
(h) {\small 40th POD basis.}
\\
\end{tabular}
\caption{\textbf{Lock exchange.} The figure shows some of the first 40 bases functions of the problem.}
\label{fig:lock_basis}
\end{figure}

Figure \ref{fig:sing} displays the decreasing singular values of the problem calculated with quantum method. The first 5 singular values exhibit rapid decay, indicating that the dominant basis functions corresponding to these singular values capture the primary energy characteristics of the original dynamical system. Figure \ref{fig:lock_basis} illustrates a subset of the first 40 basis functions calculated with quantum method, where the initial few characterize macroscopic velocity patterns, while higher-order functions resolve fine-grained flow structures. For model construction, 7 and 15 basis functions are selected to build the NIROM using both classical DKL and quantum DKL frameworks.

\begin{figure}[htbp!]
\centering
\begin{tabular}{cc}
\begin{minipage}{0.45\linewidth}
\includegraphics[width = \linewidth,angle=0,clip=true]{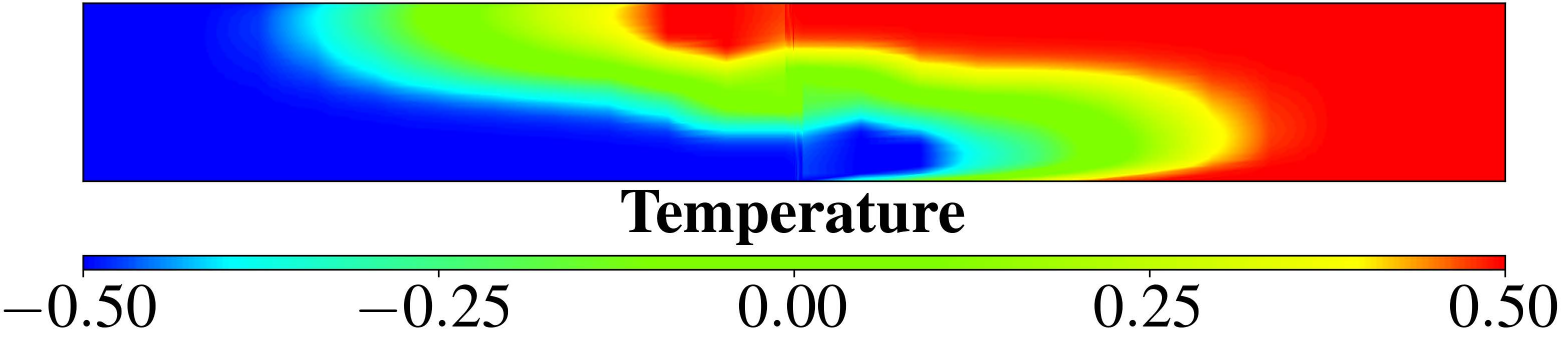}  
\end{minipage} 
&
\begin{minipage}{0.45\linewidth}
\includegraphics[width = \linewidth,angle=0,clip=true]{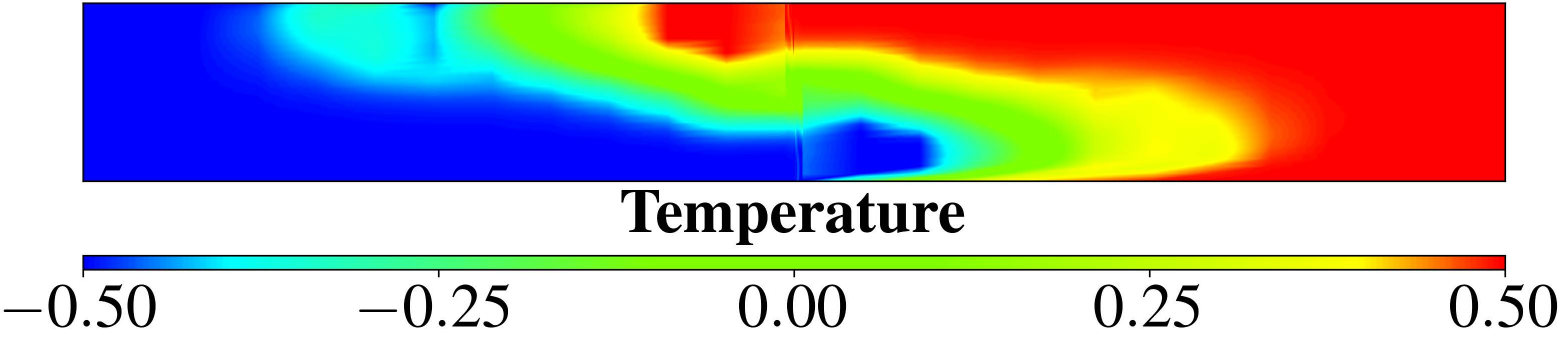}  
\end{minipage} 
 \\
(a) {\small Fully quantum framework, 7 POD bases}&
(b) {\small Classical framework, 7 POD bases}
\\

\begin{minipage}{0.45\linewidth}
\includegraphics[width = \linewidth,angle=0,clip=true]{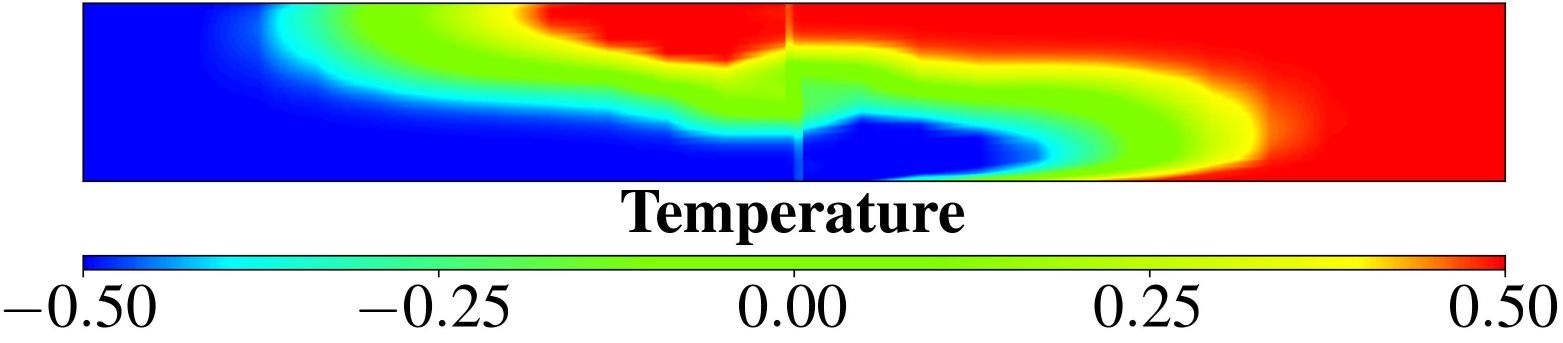}  
\end{minipage} 
&
\begin{minipage}{0.45\linewidth}
\includegraphics[width = \linewidth,angle=0,clip=true]{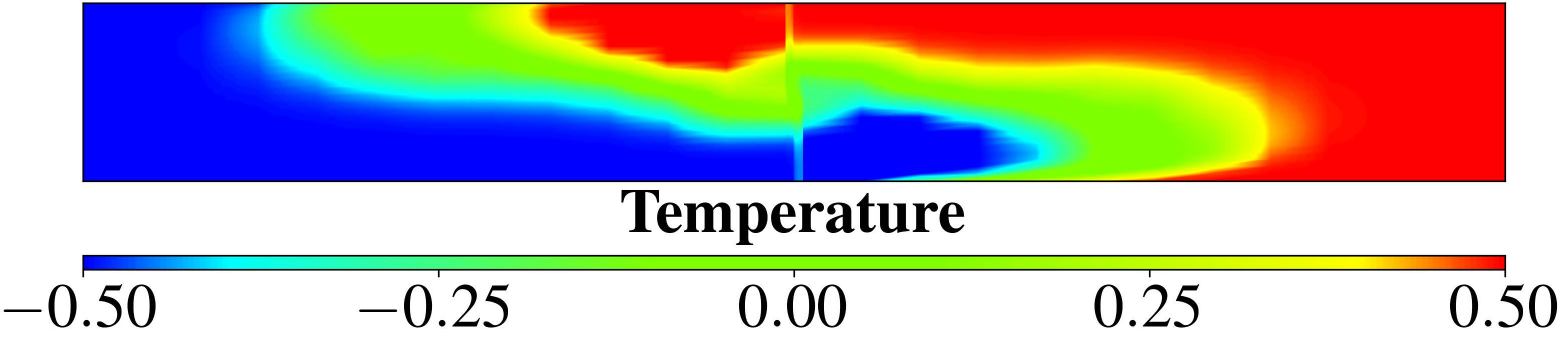}  
\end{minipage} 
 \\
(c) {\small Fully quantum framework, 15 POD bases}&
(d) {\small classical framework, 7 POD bases}
\\
\begin{minipage}{0.45\linewidth}
\includegraphics[width = \linewidth,angle=0,clip=true]{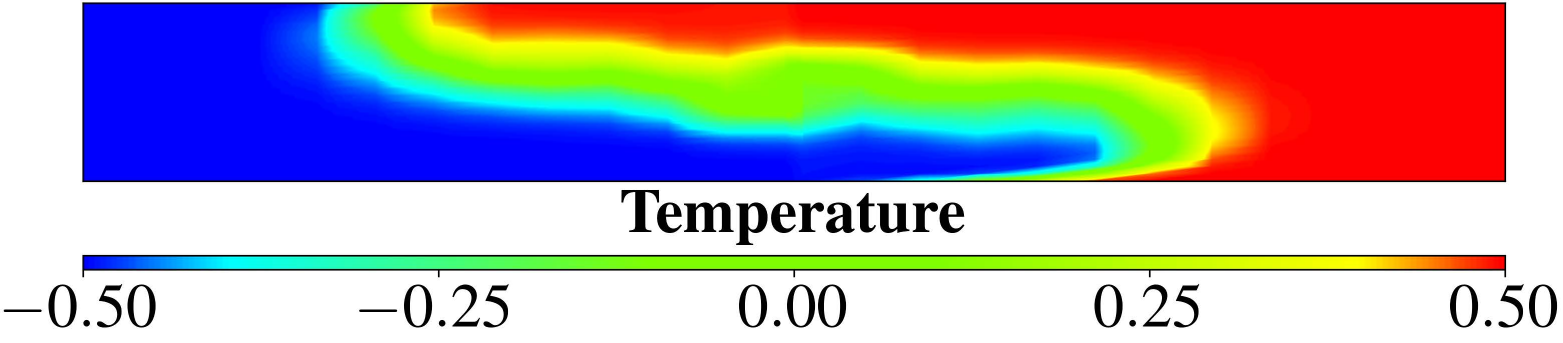}  
\end{minipage} 
&
\,
 \\
(e) {\small Full model.}&
\,
\\
\end{tabular}
\caption{\textbf{Lock exchange.} The figures displayed above show the temperature from the high fidelity model and the NIROM using classical and quantum DKL with 7 and 15 POD bases functions at time instances $t=18.36$.}
\label{fig:lock_sol}
\end{figure}

\begin{figure}[htbp!]
\centering
\begin{tabular}{cc}
\begin{minipage}{0.45\linewidth}
\includegraphics[width = \linewidth,angle=0,clip=true]{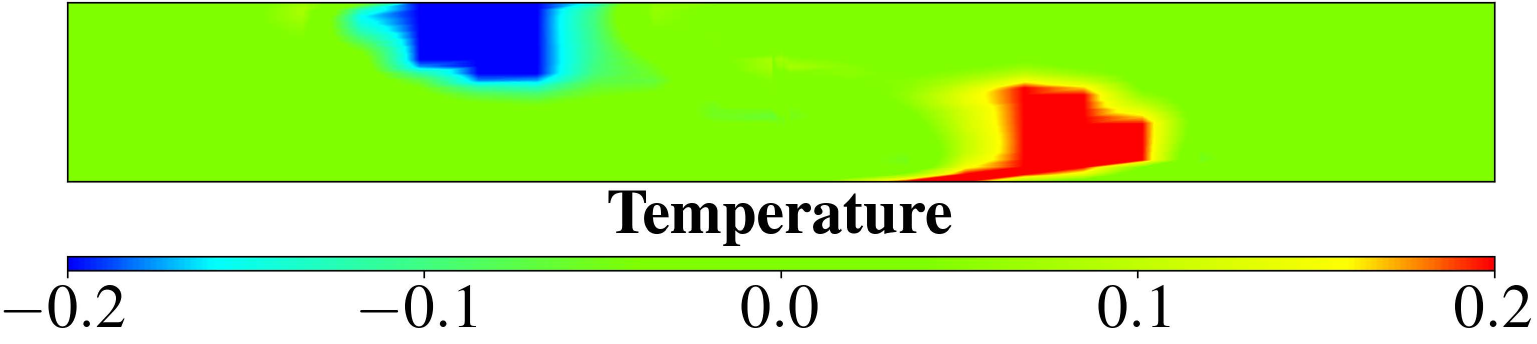}  
\end{minipage} 
&
\begin{minipage}{0.45\linewidth}
\includegraphics[width = \linewidth,angle=0,clip=true]{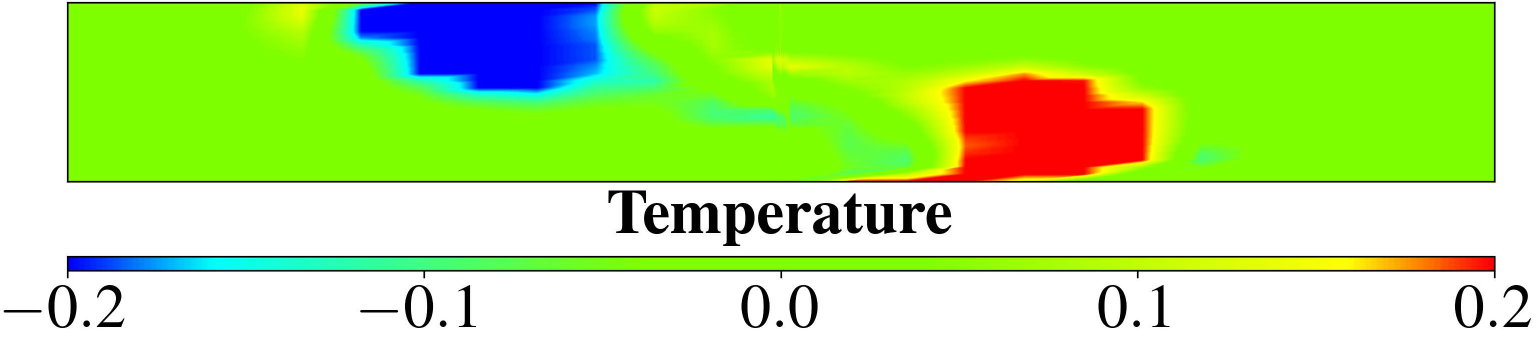}  
\end{minipage} 
 \\
(a) {\small Fully quantum framework, 7 POD bases}&
(b) {\small Classical framework, 7 POD bases}
\\

\begin{minipage}{0.45\linewidth}
\includegraphics[width = \linewidth,angle=0,clip=true]{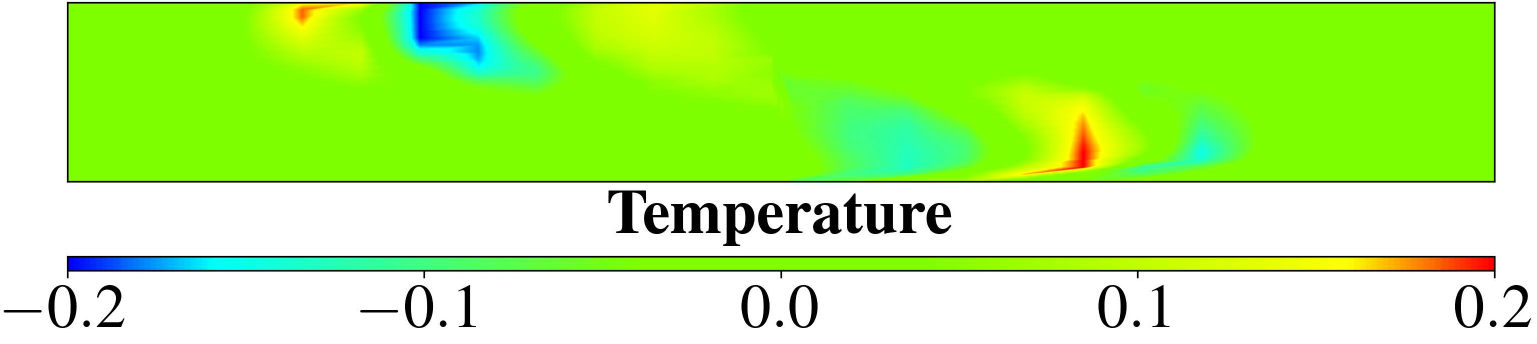}  
\end{minipage} 
&
\begin{minipage}{0.45\linewidth}
\includegraphics[width = \linewidth,angle=0,clip=true]{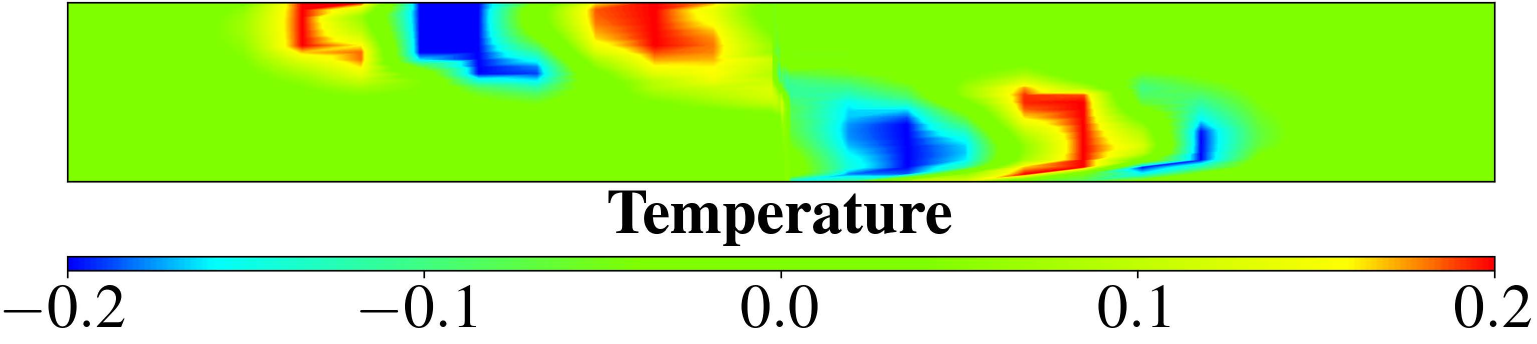}  
\end{minipage} 
 \\
(c) {\small Fully quantum framework, 15 POD bases}&
(d) {\small Classical framework, 15 POD bases}
\\
\end{tabular}
\caption{\textbf{Lock exchange at $\boldsymbol{Re=4000}$.} The figures show the temperature error between high fidelity model and NIROM with 7 and 15 POD bases at time instances $t=18.36$.}
\label{fig:lock_err}
\end{figure}

\begin{figure}[htbp!]
\centering
\begin{tabular}{c}
\begin{minipage}{0.6\linewidth}
\includegraphics[width = \linewidth,angle=0,clip=true]{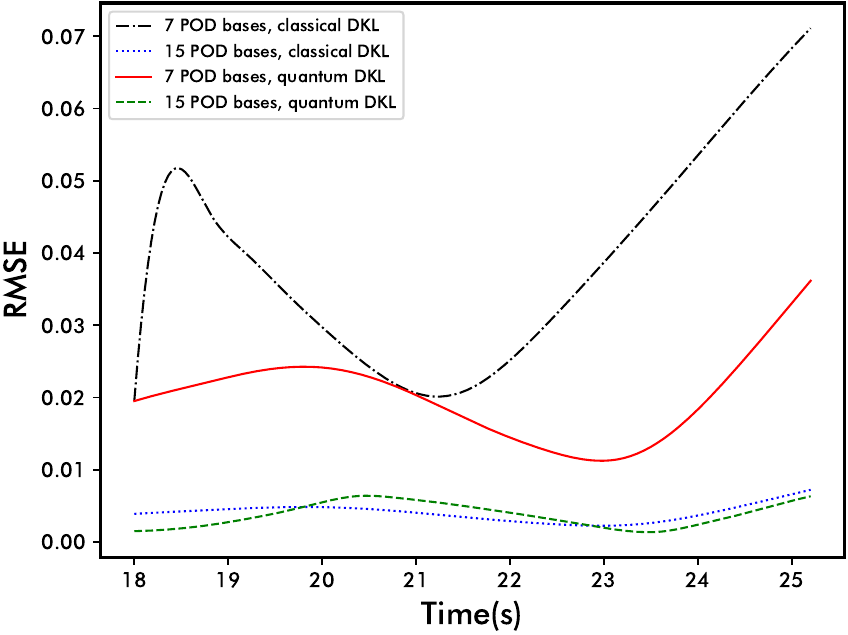}
\end{minipage} 
\\
\begin{minipage}{0.6\linewidth}
\includegraphics[width = \linewidth,angle=0,clip=true]{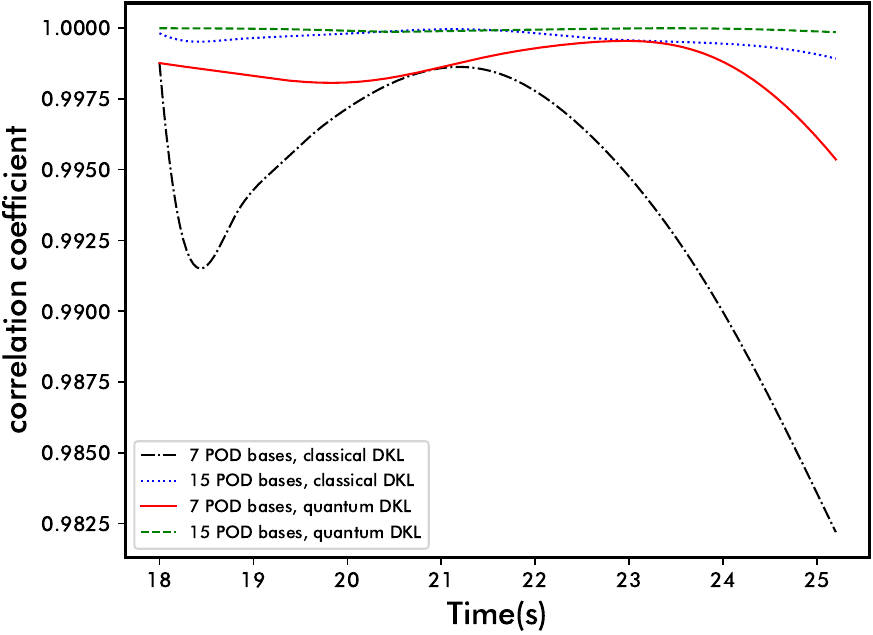}
\end{minipage} 
\end{tabular}
\caption{\textbf{Lock exchange at $\boldsymbol{Re=4000}$.} The figure shows the root mean squared error (RMSE) and correlation coefficient between the high fidelity model and NIROMs with 7 and 15 POD basis functions.}
\label{fig:rmse_corr}
\end{figure}

Figure \ref{fig:lock_sol} compares temperature solutions from the high-fidelity model, classical framework, and fully quantum framework with 7 and 15 basis functions. Solutions from the quantum framework show negligible visual discrepancies compared to the high-fidelity benchmark. Increasing the number of basis functions enhances the approximation accuracy of the NIROM. With 15 basis functions, the quantum framework achieves near-perfect alignment with the high-fidelity solution. Figure \ref{fig:lock_err} visualizes solution differences between the high-fidelity model and NIROMs. The error for the 15-basis quantum framework is visually imperceptible. Notably, the fully quantum architecture outperforms classical framework in error minimization, underscoring the quantum approach's superiority. Root mean square error (RMSE) and correlation coefficients (Figure \ref{fig:rmse_corr}) confirm that the quantum framework with 15 basis functions stabilizes with RMSE approaching 0 and correlation coefficients approaching 1. This indicates that the quantum framework captures $99.9\%$ of the original system's energy.

\section{Discussion}\label{sec:discussion}
In conclusion, this work pioneers the first integration of QML with  POD and DKL, establishing the groundbreaking QPOD-QDKL framework for NIROM of turbulent flows. This represents a significant departure from conventional NIROMs, which rely solely on classical machine learning techniques like autoencoders and GPR for flow field prediction. Our quantum-enhanced architecture introduces two key transformative innovations that fundamentally address limitations in classical turbulence modeling.

Moving beyond classical POD methods constrained by intrusive solver requirements or massive snapshot datasets, our novel Quantum POD (QPOD) leverages parameterized quantum circuits to inherently generate orthogonal basis functions through quantum eigenvalue decomposition. This innovative approach harnesses quantum parallelism to dramatically accelerate the creation of high-fidelity modal representations, a capability uniquely advantageous for resolving the complex, nonlinear vortex dynamics characteristic of turbulence, especially within low-rank regimes where classical linear subspaces struggle.

To overcome the inherent limitations of classical kernel methods in capturing intricate nonlinear features, particularly under data constraints, we introduce Quantum Deep Kernel Learning (QDKL). This novel methodology integrates core quantum phenomena, entanglement and nonlinear quantum gate operations, directly into the deep kernel learning architecture. This exploits the superior expressive power of quantum systems to significantly enhance the representational capacity of the Matern kernel, enabling it to model the complex energy transfer and multi-scale interactions in turbulent flows more effectively than classical kernels alone.

The combined power of these quantum innovations has been rigorously validated on canonical turbulent flow benchmarks. Crucially, our fully quantum-enhanced POD-DKL framework achieves prediction accuracy matching classical DKL models trained for 3000 epochs, using only 300 training epochs. This demonstrates a tenfold improvement in computational efficiency, showcasing the substantial acceleration potential unlocked by the quantum paradigm.

While promising, the current methodology faces challenges inherent to the NISQ era. Key limitations include hardware dependency, where quantum noise and limited qubit counts currently restrict experimental validation to lower-dimensional regimes, necessitating classical support for high-fidelity flows. Computational overhead also requires attention, as optimizing hybrid classical-quantum architectures and parameter tuning remain intensive processes, especially for large-scale problems.

Future advancements will focus on developing lightweight quantum encoding strategies to mitigate qubit resource demands while preserving resolution, and exploring quantum-classical heterogeneous computing frameworks to enhance parallel efficiency through task-specific hardware orchestration. This study pioneers a scalable quantum-enhanced NIROM paradigm for computational fluid dynamics, demonstrating that synergistic quantum-classical integration can overcome fundamental bottlenecks in modeling nonlinear Navier-Stokes systems. As quantum hardware matures towards fault-tolerant architectures with scalable connectivity, such hybrid frameworks hold transformative potential for enabling real-time predictive capabilities in complex industrial flow scenarios, including transient vortex dynamics and multi-physics interactions, bridging the gap between quantum algorithmic promise and practical fluid flow engineering.
\section{Method}\label{sec:method}
\subsection{Numerical experimental design}

This study employs two classical fluid mechanics benchmark problems lid-driven cavity flow, flow past a cylinder and lock exchange as validation cases. High-fidelity datasets are generated using the finite-element full-order model (FOM) in the open-source CFD software Fluidity. For the training of the quantum-enhanced DKL architecture, the Adam optimizer is adopted with a 4-layer quantum circuit, 300 training epochs, and dynamically adjusted learning rates ranging from $10^{-3}$ to $10^{-1}$. To systematically evaluate algorithmic performance, fully classical POD-DKL framework and fully quantum POD-DKL framework are compared.

\subsection{The VQC employed in QDKL}
\begin{figure}[ht]
\centering
\includegraphics[width = \linewidth,angle=0,clip=true]{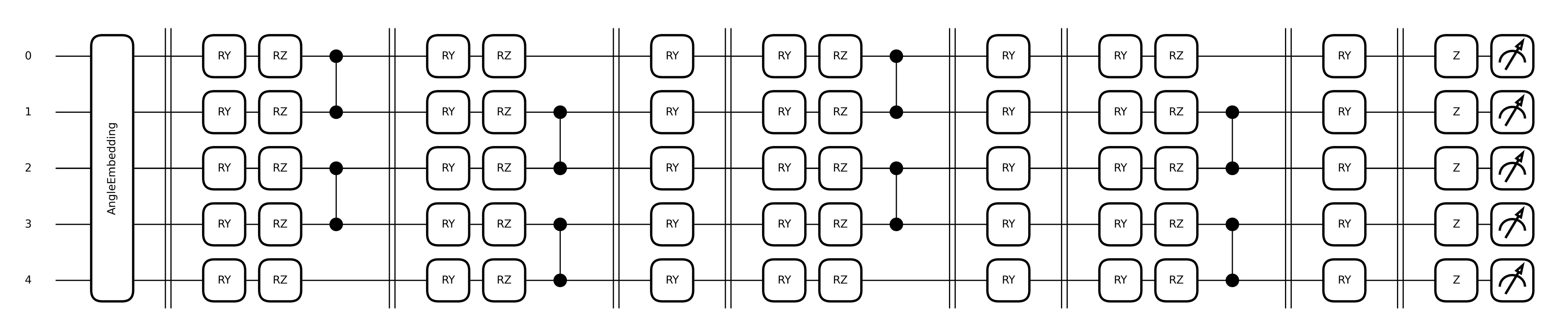} 
\caption{\textbf{A VQC usable in this algorithm.}}\label{fig: vqc_qdpl}
\end{figure}

The VQC employed consists of three components: an input encoding layer, parameterized quantum layers, and a measurement layer, with the detailed architecture shown in Figure \ref{fig: vqc_qdpl}. First, the input layer utilizes the \textit{AngleEmbedding} method to map classical input data into quantum state space. The parameterized quantum layers are constructed by cascading multiple trainable layers, where each layer contains single-qubit rotation units, an entanglement layer, and an enhanced rotation unit. Finally, Pauli-Z measurements are performed on all qubits to output a classical feature vector.

\bibliographystyle{elsarticle-num} 
\bibliography{sn-bibliography}
\end{document}